\documentclass[11pt,a4paper]{article}
\usepackage{jcappub}
\usepackage{comment}
\usepackage[utf8]{inputenc} 
\usepackage{amsfonts,amssymb, mathrsfs, amsmath, esint,bm,enumitem}
\allowdisplaybreaks

\usepackage{subcaption}
\usepackage{ragged2e}
\DeclareCaptionJustification{justified}{\justifying}
\usepackage{diagbox}
\usepackage{latexsym}
\usepackage{graphicx}
\usepackage[dvipsnames]{xcolor}
\usepackage{booktabs,multirow}
\usepackage{titlesec} 
\usepackage{physics}
\usepackage[normalem]{ulem}

\usepackage{tikz}
\usetikzlibrary{positioning,calc}

\usepackage{hyperref}
\hypersetup{colorlinks, citecolor=ForestGreen, linkcolor=BrickRed, urlcolor=RedViolet}

\usepackage[capitalize,noabbrev]{cleveref}
\crefname{equation}{Eq.}{Eqs.}
\usepackage{float}
\usepackage{placeins}
\usepackage{soul}

\newcommand{\CMBP}{\textbf{CMB-P}}
\newcommand{\CMBPA}{\textbf{CMB-PA}}
\newcommand{\CMBPS}{\textbf{CMB-PS}}
\newcommand{\CMBPAS}{\textbf{CMB-PAS}}
\newcommand{\woee}{(\sout{EE})}
\newcommand{\bao}{\textbf{B}}
\newcommand{\pantheon}{\textbf{SN}}
\newcommand{\DES}{\textbf{SN$_\text{DES}$}}
\newcommand{\union}{\textbf{SN$_\text{Union3}$}}

\newcommand{\Covmat}{\mathbf{\Sigma}}
\newcommand{\Invcovmat}{\mathbf{\Omega}}
\newcommand{\covmat}{\Sigma}

\newcommand{\Covdiag}{\mathbf{\Sigma_{\sqrt{diag}}}}
\newcommand{\Invcovdiag}{\mathbf{\Omega_{\sqrt{diag}}}}
\newcommand{\Pearsonmat}{\mathbf{R}}
\newcommand{\pearson}{R}
\newcommand{\Partialmat}{\mathbf{\rho}}

\newcommand{\taureio}{\tau_{\rm reio}}
\newcommand{\logA}{\log (10^{10}A_s)}
\newcommand{\omegacdm}{\omega_{\rm cdm}}
\newcommand{\Omegacdm}{\Omega_{\rm cdm}}
\newcommand{\Omegam}{\Omega_{\rm m}}

\newcommand{\beq}{\begin{equation}}
\newcommand{\eeq}{\end{equation}}

\title{Cosmic $\tau$ensions Indirectly Correlate with Reionization Optical Depth}

\author[a]{Itamar J.~Allali,}
\author[a,b]{Lingfeng Li,}
\author[a]{Praniti Singh,}
\author[a,c]{JiJi Fan}

\affiliation[a]{Department of Physics, Brown University, Providence, RI 02912, USA}
\affiliation[b]{International Center of Theoretical Physics-Asia Pacific, Beijing 100190, China}
\affiliation[c]{Brown Theoretical Physics Center, Brown University, Providence, RI 02912, USA}

\emailAdd{itamar\_allali@brown.edu}
\emailAdd{praniti\_singh@brown.edu}
\emailAdd{jiji\_fan@brown.edu}
\emailAdd{l.f.li165@gmail.com}

\abstract{The reionization optical depth $\tau_{\rm reio}$ has interesting connections to existing cosmological anomalies. As first studied in the context of the Hubble tension in our previous paper, a larger $\tau_{\rm reio}$, which could be achieved by removing the Planck low-$\ell$ polarization data, could boost $H_0$ slightly, resulting in a mild reduction of the tension between the early- and late-universe determinations of $H_0$. It has been shown later that a larger $\tau_{\rm reio}$ could also relieve other anomalies including: the tension between BAO and CMB data, the neutrino mass tension, and the latest DESI plus supernovae data's tension with the standard cosmological constant scenario. In this paper, we systematically analyze the correlations between $\tau_{\rm reio}$ and relevant cosmological parameters in the existing cosmic observation anomalies. In addition to Pearson correlation coefficients extracted directly from the covariance matrix, we also study partial correlation coefficients which measure intrinsic relationships between pairs of parameters removing the influence of other parameters. 
We show that $\tau_{\rm reio}$ has weak intrinsic correlations with the parameters responsible for the tensions and anomalies discussed. The large direct Pearson correlations that 
allow larger $\tau_{\rm reio}$ inferences to alleviate the cosmological tensions each arise
 from complicated networks through multiple parameters. 
 As a result, the relationships between $\tau_{\rm reio}$ and each anomaly are not independent of each other.
We also employ our method of computing correlations to clarify the impact of large scale polarization data, and comment also on the effects of CMB observations from ACT and SPT. 
}

\begin{document}
\maketitle
\vspace{1em}\noindent

\section{Introduction}
\label{sec:intro}

The $\Lambda$CDM model is the most widely accepted cosmological model, compatible with a variety of observations spanning many epochs of cosmic history. Within this six-parameter framework, one such parameter is the reionization optical depth, $\taureio$. The era of reionization marks a key phase in cosmic history, signaling the end of the “dark ages,” when the universe was mostly filled with neutral gas, and the beginning of star and galaxy formation accompanied by the reionization of baryonic matter.

Currently, $\taureio$, is constrained through two different observations: measurement of the Cosmic Microwave Background (CMB) and astrophysical observations of early luminous galaxies. The CMB provides a probe of $\taureio$ by measuring CMB photons scattered by free electrons produced during reionization.  This scattering suppresses the temperature anisotropy power spectrum at small angular scales and enhances the large-scale E-mode polarization signal, allowing precise constraints on $\taureio$ (see \cite{Dodelson:2003ft} for further details). On the other hand, the astrophysical approach \cite{Robertson:2015uda, McQuinn:2015icp} indirectly infers $\taureio$ by studying the population of high redshift luminous galaxies, their production rate of ionizing photons, and the fraction of these photons that escape into the intergalactic medium to ionize neutral hydrogen. Although the \textit{Planck} satellite CMB measurements currently give $\taureio=0.0544 \pm 0.0073$ \cite{Planck:2018vyg}, recent astrophysical studies (or galaxy surveys) from the James Webb Space Telescope (JWST) \cite{Jakobsen_2022,2023PASP..135d8003W} suggest a higher value, potentially $\taureio\geq0.07$ \cite{Munoz:2024fas}, mainly driven by a higher photon production rate at early times \cite{simmonds2023,Endsley2024} and, sub-dominantly, the observation of more star-forming galaxies at high redshift \cite{2024ApJ...969L...2F, Finkelstein_2023}. Moreover, several recent low-redshift surveys \cite{Chisholm_2022} suggest a higher photon escape fraction, which, when extrapolated to higher redshifts, implies an even greater value of $\taureio \sim 0.096$.

The JWST studies are potentially subject to substantial astrophysical uncertainties from observational bias in galaxy surveys, parameter extrapolation, and baryon physics modeling (see also \cite{Mukherjee:2024cfq,Paoletti:2024lji, Zhu:2024xrt,Cain:2024fbi}). Meanwhile, the precise constraints on $\taureio$ using CMB data is mainly driven by large-scale (low-$\ell$) E-mode polarization data which are subject to largest uncertainties among all the CMB measurements. Interestingly, Ref. \cite{Giare:2023ejv} has shown that the exclusion of large-scale CMB data pushes the value of $\taureio$ to a higher value, in the range preferred by the JWST studies.

While these new observations potentially challenge our current understanding of reionization, it is crucial to explore their impact on existing cosmological tensions/anomalies. In this work, we will focus on:

\begin{itemize}

    \item {\bf The Hubble tension:} This refers to the discrepancy in the value of the current expansion rate, $H_0$, inferred from CMB measurements versus more direct observations of supernovae-based distance ladder methods \cite{Scolnic:2023pga, Riess:2021jrx,Breuval:2024cqf,Freedman:2021ahq,Freedman:2023zos,Freedman:2024qnm,Wong:2019kwg,Riess:2024xlm}. 

    \item {\bf The BAO-CMB tension:} There is a mild discrepancy between CMB datasets and the recent baryon acoustic oscillations (BAO) observations from the Dark Energy Spectroscopic Instrument (DESI) \cite{DESI:2025zgx} in the most directly inferred parameters from BAO: the matter density, $\Omega_m$ and $h r_d$, with $h = H_0/(100$ km/s/Mpc$)$ and $r_d$ the physical size of the sound horizon at the baryon drag epoch.
    
    \item {\bf Neutrino mass:} There is a notable tension regarding the sum of neutrino masses $\sum m_\nu$ inferred from cosmological measurements and the lower bound imposed by neutrino oscillation experiments ($\sum m_\nu \geq 0.06$ eV for normal ordering and 0.1 eV for inverted hierarchy \cite{ParticleDataGroup:2024cfk}). Within the $\Lambda$CDM framework, recent BAO measurements from DESI, when combined with the Planck CMB data, yield a posterior for $\sum m_\nu$ peaking near zero (for prior $\sum m_\nu \geq 0$), giving rise to an upper bound of $\sum m_\nu < 0.0642$ eV (95\% C.L.)\cite{DESI:2025ejh} and seeming to rule out inverted ordering.\footnote{It has been shown that the data even favor unphysical negative extrapolations for the neutrino mass sum, see e.g. \cite{DESI:2025ejh,Craig:2024tky,Green:2024xbb,Graham:2025fdt}.
    }

    \item {\bf Dynamical Dark Energy:} Recent BAO measurements from DESI, combined with CMB data (and various supernovae datasets), show a $\sim2-4 \sigma$ preference for dynamical dark energy (DE) over the standard $\Lambda$CDM model, when employing the Chevallier-Polarski-Linder (CPL) parameterization \cite{Chevallier:2000qy,Linder:2002et} of the DE equation of state: $w(a)=w_0+w_a(1-a)$, which we refer to as the $w_0w_a$CDM \cite{DESI:2025zgx} model.
\end{itemize}

Recent studies have explored the impact of $\taureio$ measurements on these existing cosmological tensions. For instance, in our previous paper \cite{Allali:2025wwi}, it was first pointed out that increasing $\taureio$ by excluding low-$\ell$ EE CMB data can mildly alleviate the Hubble tension. Similarly, Ref. \cite{Jhaveri:2025neg, Sailer:2025lxj} showed that raising $\taureio$ to around $0.09$ relaxes the constraint on neutrino mass and reduces the preference for evolving DE. Additionally, Ref. \cite{Sailer:2025lxj} suggested that the BAO-CMB tension can also be mitigated by an increase in $\taureio$. Other studies in the literature have recently focused on probing the epoch of reionization, see \cite{Cheng:2025cmb,Cain:2025usc,Namikawa:2025doa,Tan:2025cua}.

Motivated by these findings, in this work we investigate the correlations between $\taureio$ and other cosmological parameters within the context of the tensions described above. Specifically, we study both the Pearson correlation coefficients between parameters of interest, as well as partial correlation coefficients which take into account the effects of other parameters. 
We employ these measures of correlation to glean information about the relationship between $\taureio$ and each tension/anomaly.
We begin in \cref{sec:data} by describing the datasets and methodology used in this work. In \cref{sec:results_all}, we systematically present the correlations of $\taureio$ with other parameters within the context of all four tensions studied, using CMB data from \textit{Planck}, BAO from DESI, and supernovae measurements from Pantheon+. In \cref{sec: diff data}, we examine the effects of including or excluding different combinations of datasets, including CMB data from Atacama Cosmology Telescope (ACT) and the South Pole Telescope (SPT), as well as other supernovae datasets. Finally, we give concluding remarks in \cref{sec:conclusion}.

\section{Methodology and Datasets}
\label{sec:data}
We conduct Markov Chain Monte Carlo (MCMC) simulations to explore the parameter space of several cosmological models and assess the correlations between parameters of interest. In particular, we will assess both the Pearson correlations and the partial (intrinsic) correlations between sets of parameters, as discussed further below, in the context of several combinations of datasets. We consider the following CMB observations:
\begin{itemize}
    \item \CMBP: CMB observations from the {\it Planck} satellite, including: the Public Release 3 (PR3) {\tt Commander} likelihood for the low-$\ell$ ($\ell < 30$) TT power spectrum \cite{Planck:2019nip}; the PR3 {\tt SimAll} likelihood for the low-$\ell$ polarization EE spectrum \cite{Planck:2019nip}; the PR4 {\tt Camspec} likelihood for high-$\ell$ TT, TE, and EE spectra ($30<\ell<2500$ TT, $30<\ell<2000$ TE, EE) \cite{Efstathiou_2021,Rosenberg_2022,Planck:2020olo}; and the combined {\it Planck} PR4 \cite{Carron:2022eyg} and ACT Data Release 6 (DR6) \cite{ACT:2023dou,ACT:2023kun} lensing likelihood.
    \item \CMBPA: The same as \CMBP, with the {\it Planck} high-$\ell$ data restricted to TT $\ell \in [30,1800]$, TE $\ell \in [30,750]$, EE $\ell \in [30,750]$; and the addition of the ACT DR6 multi-frequency likelihood {\tt MFLike} \cite{ACT:2025fju} for high-$\ell$ TT $\ell \in [1800,8500]$, TE $\ell \in [750,8500]$, EE $\ell \in [750,8500]$.
    \item \CMBPS: The same as \CMBP, with the {\it Planck} high-$\ell$ data restricted to TT $\ell \in [30,2000]$, TE $\ell \in [30,1200]$, EE $\ell \in [30,1200]$; and the addition of the SPT-3G D1 likelihood {\tt candl} \cite{SPT-3G:2025bzu} for high-$\ell$ TT $\ell \in [1200,3000]$, TE $\ell \in [1200,4000]$, EE $\ell \in [1200,4000]$.\footnote{In the combinations \CMBPS~and \CMBPAS, there is some overlap between the SPT and {\it Planck} data in the TT channel for our choices of $\ell$; since the sky fraction of SPT is small, we can approximate these observations to be uncorrelated as done in \cite{SPT-3G:2025bzu}.}
    \item \CMBPAS: The same as \CMBP, with the {\it Planck} high-$\ell$ data restricted to TT $\ell \in [30,1800]$, TE $\ell \in [30,750]$, EE $\ell \in [30,750]$; the addition of the ACT DR6 multi-frequency likelihood {\tt MFLike} \cite{ACT:2025fju} for high-$\ell$ TT $\ell \in [1800,8500]$, TE $\ell \in [750,8500]$, EE $\ell \in [750,8500]$; and the SPT-3G D1 likelihood {\tt candl} \cite{SPT-3G:2025bzu} for high-$\ell$ TT $\ell \in [1200,3000]$, TE $\ell \in [1200,4000]$, EE $\ell \in [1200,4000]$.\footnote{In the combination \CMBPAS, there is also an overlap between ACT and SPT data, which again can be taken to be uncorrelated as done in \cite{SPT-3G:2025bzu}, since the sky fraction overlap between the two experiments is small. }
    \item \{\CMBP\woee, \CMBPA\woee, \CMBPS\woee, \CMBPAS\woee\}: The same as the four datasets above with the removal of the low-$\ell$ EE {\it Planck} data from the {\tt SimAll} likelihood.
\end{itemize}
When combining multiple CMB datasets, we have chosen the ranges of $\ell$ for each experiment such that {\it Planck} is chosen where it achieves higher precision, and SPT/ACT are chosen when their precision exceeds {\it Planck}, according to the comparison's made in \cite{ACT:2025fju,SPT-3G:2025bzu}. In addition to CMB data, we include the following BAO and supernovae (SN) data:
\begin{itemize}
    \item \bao: BAO measurements from the DESI data release 2 (DR2) for effective redshifts $z = 0.295$, $0.51$, $0.706$,
 $0.934$, $1.321$, $1.484$, $2.33$ \cite{DESI:2025zgx}.
    \item \pantheon: Type Ia supernovae catalog from Pantheon+ \cite{Scolnic:2021amr}.
    \item \DES: Type Ia supernovae catalog for the Dark Energy Survey (DES) 5-year dataset \cite{DES:2024jxu}.
    \item \union: Type Ia supernovae catalog from the Union3 compilation \cite{Rubin:2023jdq}.
\end{itemize}
When not focusing on results that rely heavily on the choice of supernovae dataset, we will include the Pantheon+ set as default. Thus, the baseline dataset for this work will be the combination \CMBP+\bao+\pantheon~and the corresponding one without low-$\ell$ EE {\it Planck} data, \CMBP \woee+\bao+\pantheon.

We employ the Boltzmann-solver {\tt CLASS} \cite{Lesgourgues:2011re,Blas:2011rf} for computing the cosmic evolution for each point in parameter space, sampled by the {\tt Cobaya} sampler \cite{Torrado:2020dgo,2019ascl.soft10019T}. We use {\tt getdist} for analyzing the samples \cite{Lewis:2019xzd}.

\subsection{Evaluating Correlations}
\label{ssec:partialcorr}
We will assess correlations between parameters using the results of the MCMC analyses. First, the covariance matrix $\covmat_{ij}$ of a subset of parameters can be obtained straightforwardly from the MCMC samples. The diagonal components $\covmat_{ii}$ correspond to the variance of the $i^\text{th}$ parameter, and the off-diagonal $\covmat_{ij}$ correspond to the covariance between the $i^\text{th}$ and $j^\text{th}$ parameters. Thus, the correlation between two parameters can be computed as follows:
\begin{equation}
    \pearson_{ij} = \frac{\covmat_{ij}}{\sqrt{\covmat_{ii}\covmat_{jj}}}=\frac{\sigma_{ij}^2}{\sqrt{\sigma_i^2\sigma_j^2}} \, ,
\end{equation}
where $\pearson_{ij}$ is the Pearson correlation coefficient between the $i^\text{th}$ and $j^\text{th}$ parameters, corresponding to the correlation between the sets of these two parameters contained in the MCMC sample, under the assumption that these parameters are independent of other parameters. This is equivalent to the ratio of the covariance between the two parameters $\sigma_{ij}^2$ to the square root of the product of their respective variances $\sigma_i^2$ and $\sigma_j^2$. Equivalently, the correlation matrix $\Pearsonmat$ can be obtained as
\begin{equation}
    \Pearsonmat = (\Covdiag)^{-1}\times \Covmat \times (\Covdiag)^{-1} \, ,
      \label{eq:Pearson1}
\end{equation}
with $\Covmat$ the covariance matrix, and $\Covdiag$ is a diagonal matrix containing the standard deviations of each parameter, or equivalently the square roots of the diagonal components of the covariance matrix.

We wish also to assess the intrinsic relationship between parameters, where the variation in other parameters is taken into account and effectively removed. Known as {\it partial correlations}, these measures are an indication of the correlation between a pair of parameters while factoring out the influence of any other correlated parameters. For instance, in the case of three correlated random parameters $X$, $Y$ and $Z$, the partial correlation between $X$ and $Y$ is calculated by removing their correlation coefficient with $Z$:
\begin{equation}
    \rho_{XY\cdot Z} = \frac{R_{XY}-R_{XZ}R_{ZY}}{\sqrt{1-R_{XZ}^2}\sqrt{1-R_{YZ}^2}}~,
\label{eq:partial1}
\end{equation}
where $R_{ij}$ is the Pearson correlation defined in Eq.~\eqref{eq:Pearson1}. In a sample that is sufficiently Gaussian, the partial correlation between parameter $X$ and $Y$ can be considered as the Pearson correlation between them when all other parameters are fixed to their best-fit values. The approach is close to Figure 1 of~\cite{Allali:2025wwi} by doing fits with only two parameters.

Obtaining the partial correlations for multiple parameters is straightforward, similar to the Pearson correlations but working instead with the inverse of the covariance matrix, also known as the precision matrix, $\Invcovmat\equiv \Covmat^{-1}$. The matrix of partial correlations $\Partialmat$ is given by
\begin{equation}
   \Partialmat = -(\Invcovdiag)^{-1}\times \Invcovmat \times (\Invcovdiag)^{-1} \, ,
   \label{eq:partial2}
\end{equation}
where similarly $\Invcovdiag$ is a diagonal matrix containing the square roots of the diagonal elements of $\Invcovmat$.\footnote{The relation of Eq.~\eqref{eq:partial2} is mathematical and independent of the data details. However, it works best when the data is sufficiently Gaussian. In this case, the partial correlation obtained will be close to its original meaning, namely the Pearson correlation of the parameters of interest with all other parameters fixed. This condition is satisfied by the cosmological MC samples, as we will see below in \cref{sec:results_all}. } In practice, one can also only remove a subset of remaining parameters by considering a submatrix of the covariance matrix with irrelevant parameters excluded. In this work, we always ignore the nuisance parameters when calculating partial correlations, as they behave as noise to cosmological parameters. Note that the relation of Eq.~\eqref{eq:Pearson1} and~\eqref{eq:partial2} are dual to each other since $\covmat = \Invcovmat^{-1}$. In other words, one could also start with the partial correlation $\rho$ and obtain the Pearson correlation $\pearson$. In this sense, the Pearson correlation $\pearson$ and partial correlation $\Partialmat$ are dual descriptions of the data, with the latter discussed less often in cosmology.

\section{Correlations of $\taureio$ and Cosmic Tensions}
\label{sec:results_all}
In this section, we will discuss the impact of $\taureio$ on existing cosmological anomalies, listed in \cref{sec:intro}, one by one. To demonstrate the effects of increasing $\taureio$, we will compare MC samples obtained using CMB data with and without the low-$\ell$ EE data, as the removal of this data results in a larger value of $\taureio$. To avoid clumsiness, we will present results mainly using the simplest CMB dataset \CMBP\woee~or \CMBP, plus \bao~and \pantheon. We will discuss the effects of adding ACT and SPT datasets in \cref{sec: diff data}, as they will not modify the main conclusions. Additional results for correlations in the data combination \CMBPAS~are given in \cref{app:PAS_corr}, and results involving \DES~and \union~are presented in \cref{app:DES_union}. Also, though we do not focus on presenting constraints for cosmological parameters in this work, we do provide posterior statistics and plots for the novel combination of CMB data presented in this work, \CMBPAS,~in \cref{app:PAS_post}.

\subsection{Hubble Tension}\label{sec:H0tension}

\begin{figure}
    \centering
    \includegraphics[width=0.75\linewidth]{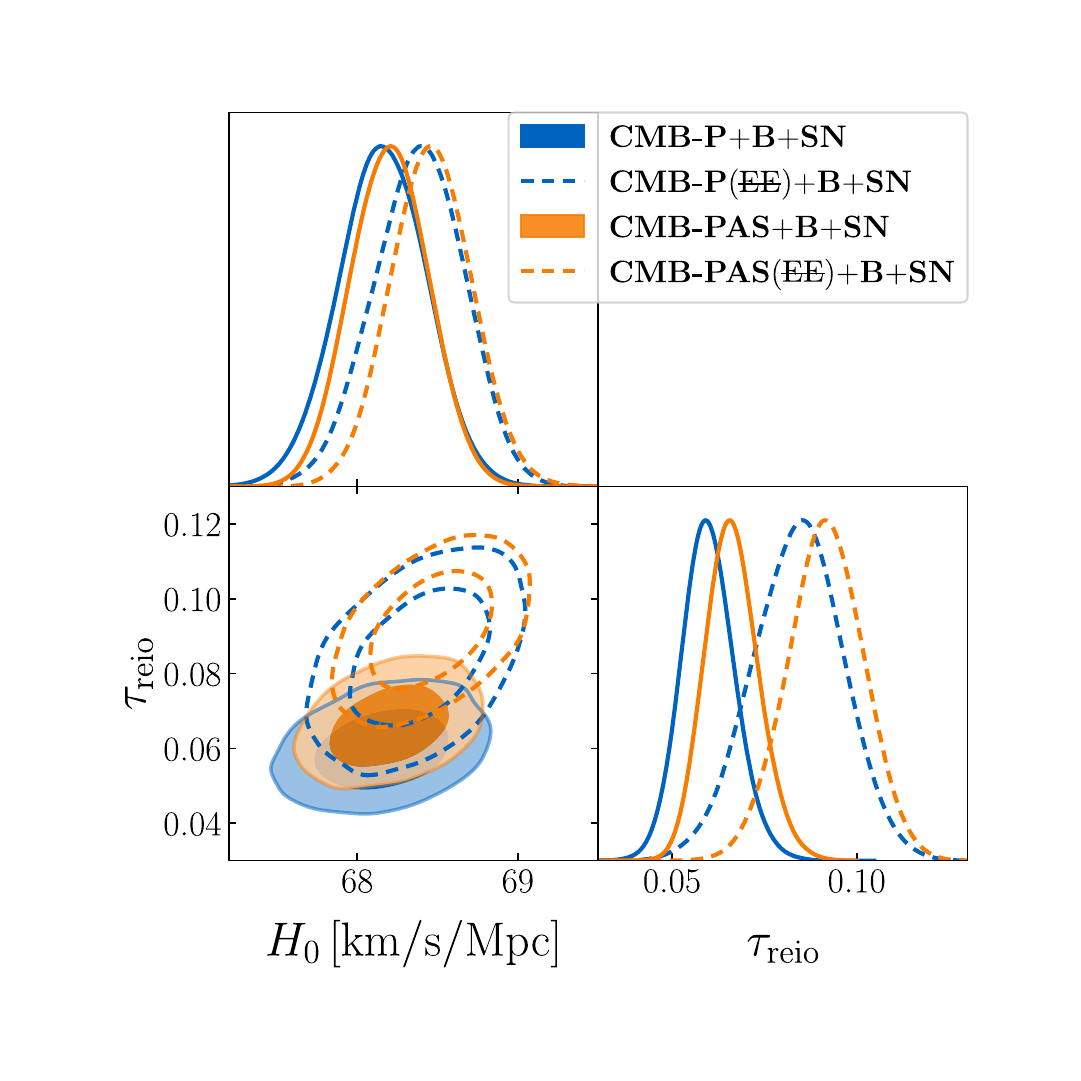}
    \caption{The one- and two-dimensional posterior distributions of $H_0$ and $\taureio$ in a series of fits to four different combinations of datasets, assuming $\Lambda$CDM. }
    \label{fig:H0_tau}
\end{figure}

\begin{table}[]
    \centering
 \begin{tabular}{lcccccc}
\toprule
{\bf Pearson} & $\logA$ & $n_s$ & $H_0$ & $\omega_b$ & $\omegacdm$ & $\taureio$ \\
\midrule
$\logA$ & 1.00 & 0.30 & 0.43 & 0.23 & -0.41 & 0.98 \\
$n_s$ & 0.30 & 1.00 & 0.49 & 0.23 & -0.48 & 0.33 \\
$H_0$ & 0.43 & 0.49 & 1.00 & 0.59 & -0.89 & \textbf{0.43} \\
$\omega_b$ & 0.23 & 0.23 & 0.59 & 1.00 & -0.25 & 0.22 \\
$\omegacdm$ & -0.41 & -0.48 & -0.89 & -0.25 & 1.00 & -0.43 \\
$\taureio$ & 0.98 & 0.33 & \textbf{0.43} & 0.22 & -0.43 & 1.00 \\
\bottomrule
\end{tabular}
\begin{tabular}{lcccccc}
\toprule
 {\bf Partial} & $\logA$ & $n_s$ & $H_0$ & $\omega_b$ & $\omegacdm$ & $\tau_\mathrm{reio}$ \\
\midrule
$\logA$ & --- & -0.13 & 0.09 & 0.00 & 0.10 & 0.98 \\
$n_s$ & -0.13 & --- & 0.11 & -0.02 & -0.05 & 0.16 \\
$H_0$ & 0.09 & 0.11 & --- & 0.80 & -0.92 & \textbf{-0.08} \\
$\omega_b$ & 0.00 & -0.02 & 0.80 & --- & 0.71 & 0.00 \\
$\omegacdm$ & 0.10 & -0.05 & -0.92 & 0.71 &--- & -0.11 \\
$\taureio$ & 0.98 & 0.16 & \textbf{-0.08} & 0.00 & -0.11 & --- \\
\bottomrule
\end{tabular}
    \caption{Pearson correlations (top) and partial correlations (bottom) between the standard six parameters in $\Lambda$CDM  with the dataset \CMBP\woee+\bao+\pantheon. The correlations between $\taureio$ and $H_0$ are highlighted in bold.}
    \label{tab:LCDMcorr}
\end{table}

\begin{figure}[h!]
    \centering
    \includegraphics[width=0.45\linewidth]{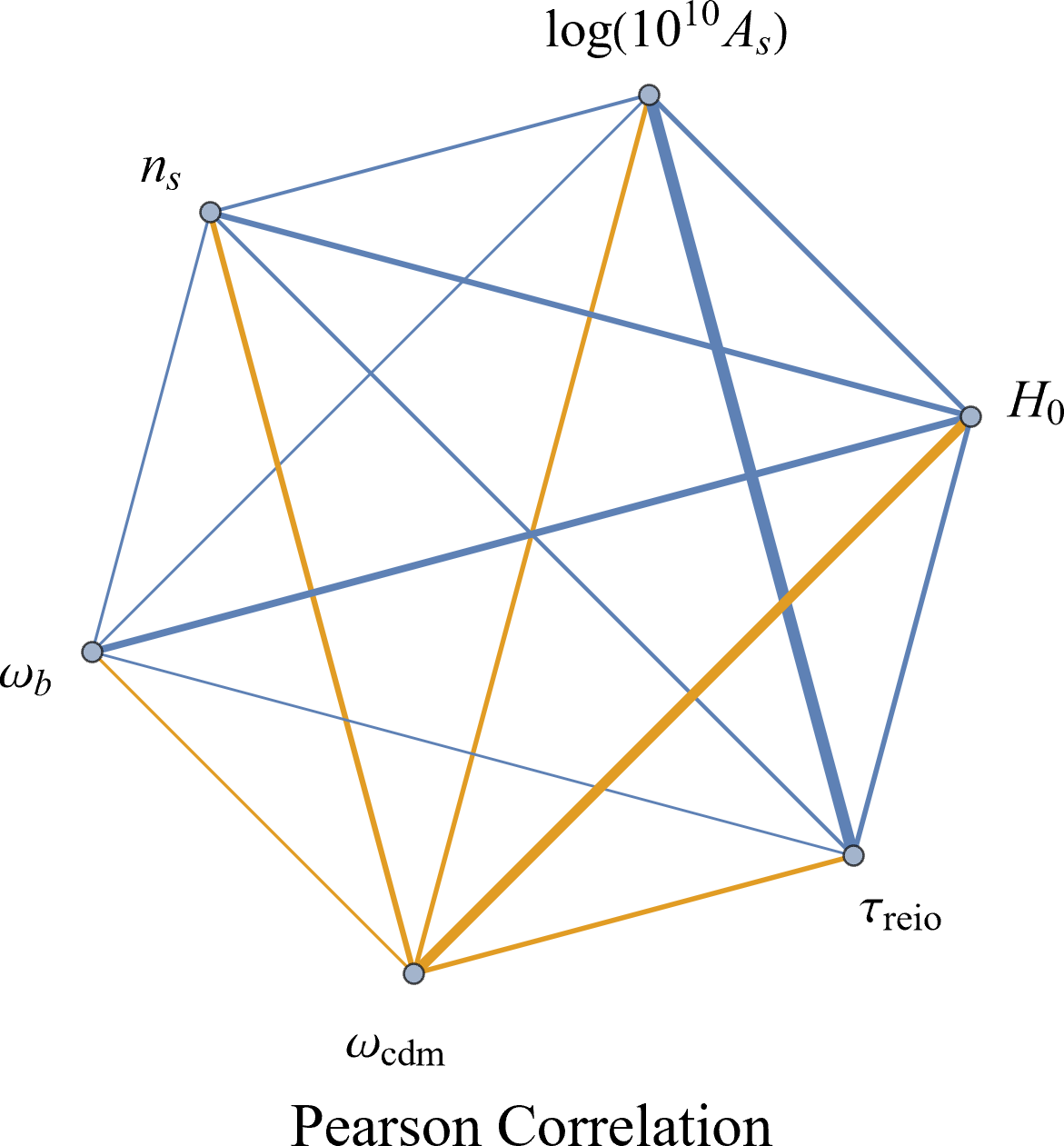}
    \includegraphics[width=0.45\linewidth]{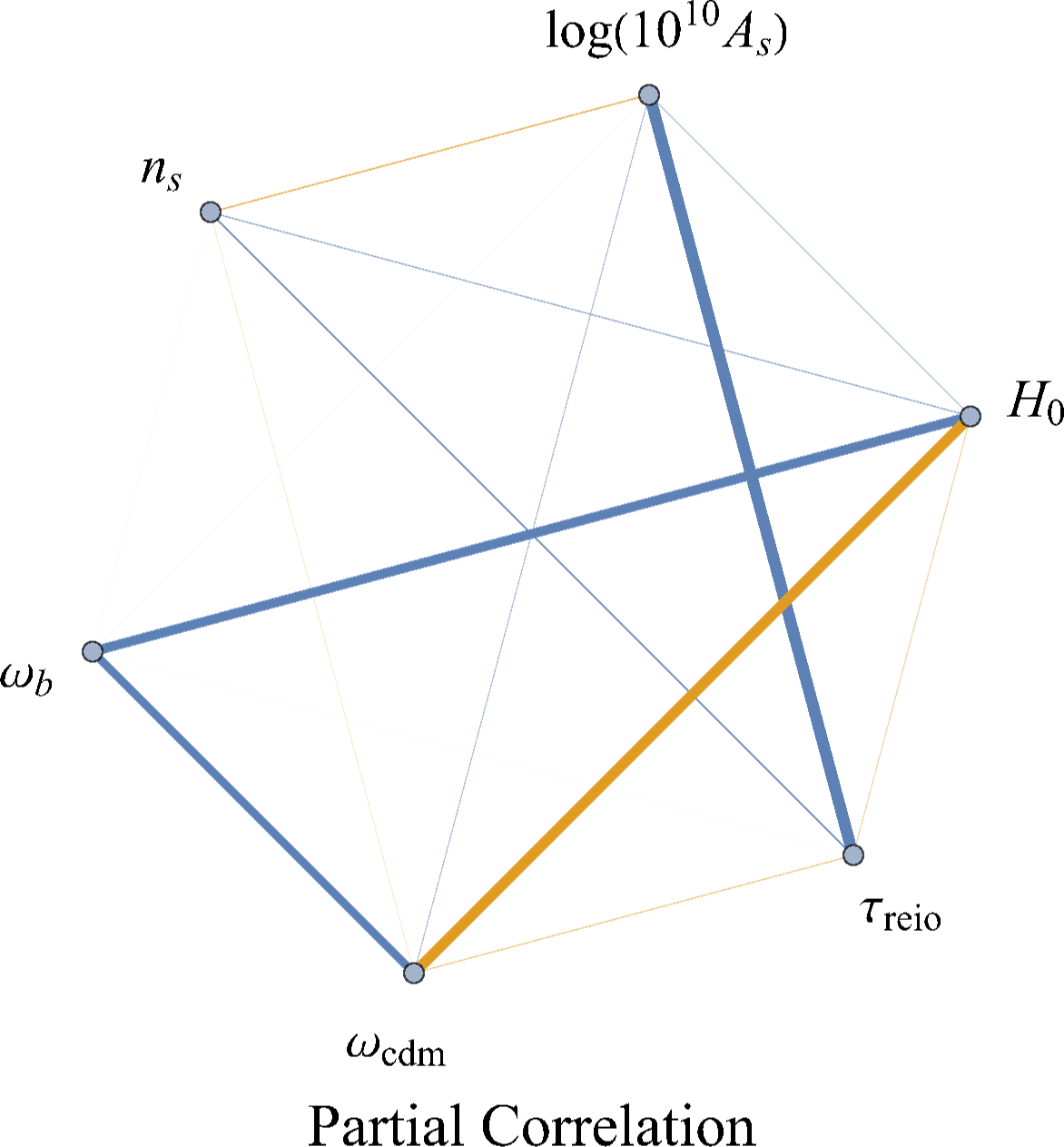}
    \caption{Pearson correlations (left) and partial correlations (right) between parameters in $\Lambda$CDM, with the dataset \CMBP\woee+\bao+\pantheon. Blue lines indicate positive correlations, orange lines indicate negative, and the line thickness is broadcast to a linear scale of the correlation strength. }
    \label{fig:corr_LCDM_1}
\end{figure}

Let us first discuss the impact of $\taureio$ on the Hubble tension. As has been observed in our previous work \cite{Allali:2025wwi}, $\taureio$ has a mild positive correlation with the Hubble constant $H_0$ in $\Lambda$CDM, as well as in some popular scenarios to relieve the Hubble tension. Here we focus on $\Lambda$CDM, with the standard set of six parameters: $A_s$ is the amplitude of primordial scalar fluctuations, $n_s$ the scalar spectral index, $H_0$ the Hubble expansion rate today, $\Omega_b h^2 \equiv \omega_b$ the baryon density today with $h=H_0/100$, $\Omegacdm h^2 \equiv \omegacdm$ the cold dark matter density today, and $\taureio$ the reionization optical depth. For $\Lambda$CDM in this work, we treat neutrinos as three degenerate mass eigenstates with total mass $\sum m_\nu =0.06$ eV (we allow for the neutrino mass sum to be a free parameter in \cref{sec:nu_mass}.) Due to the positive $\taureio$-$H_0$ correlation, with the removal of low-$\ell$ EE data and the resulting larger inferred value of $\taureio$, $H_0$ increases mildly and the significance of the Hubble tension is reduced. This can be seen from \cref{fig:H0_tau}. In the figure, for either \CMBP~or \CMBPAS, without low-$\ell$ EE data, the positive correlation between $\taureio$ and $H_0$ is more evident, together with simultaneously enhanced $\taureio$ and $H_0$ (to a lesser degree).

In order to evaluate the nature of this correlation, we compute the Pearson correlations as well as the (intrinsic) partial correlations between each pair of $\Lambda$CDM model parameters when fit to the \CMBP\woee+\bao+\pantheon~dataset\footnote{We focus on the correlation coefficients for the \CMBP\woee+\bao+\pantheon~dataset for most of this work. For \CMBP+\bao+\pantheon, the correlations remain qualitatively similar.}, reported in \cref{tab:LCDMcorr} (see \cref{app:PAS_corr} for correlations when using \CMBPAS). First, one can observe that the Pearson correlation between $\taureio$ and $H_0$ is moderate and positive, with a value of 0.43 as shown in bold. However, when computing the partial correlations, we can see that the correlation between $H_0$ and $\taureio$ more or less vanishes, with a value of $-0.08$, indicating that this is not an intrinsic relationship. Rather, the correlation must be carried through a series of correlations between $\taureio$ and other parameters.

One can also visualize these relationships through the correlation diagrams in \cref{fig:corr_LCDM_1}. Here, we depict the Pearson correlations on the left and the partial correlations on the right. Blue lines indicate positive correlations, orange lines indicate negative, and the line thickness is broadcast to a linear scale of the correlation strength. Focusing on the partial correlations on the right-hand side of \cref{fig:corr_LCDM_1}, one can draw paths from $\taureio$ to $H_0$ passing through other parameters to try to recreate the positive relationship between the two. For instance, the path $\taureio\to n_s \to H_0$
involves all positive partial correlations, and would contribute to the positive Pearson correlation between $\taureio$ and $H_0$. Similarly, a path involving $\taureio\to \omegacdm\to H_0$ involves two negative partial correlations and would thus also contribute positively to the overall relationship between $H_0$ and $\taureio$. Notably, some paths in the correlation-space will also contribute negatively to the correlation between $H_0$ and $\taureio$, e.g. the path $\taureio \to \omegacdm\to \omega_b \to H_0$.\footnote{Interestingly, the tight Pearson correlation between $\logA$ and $\taureio$ suggests that partial correlations of $\taureio$ with any parameter $X$ should be very small and opposite to the correlation of $\logA$ with $X$ (mostly canceling in \cref{eq:partial1}), as evident in \cref{tab:LCDMcorr}.} Ultimately, paths with stronger correlations will contribute more significantly, resulting in the overall positive Pearson correlation between $\taureio$ and $H_0$. 

Using the partial correlation coefficients and the correlation diagrams, then, one can clearly see that the correlation of $\taureio$ and $H_0$ within the data considered involves many parameters and is not an intrinsic characteristic of these parameters. Nonetheless, a larger inferred value of $\taureio$ still results in a shift in $H_0$ and a mild reduction of the Hubble tension.

\subsection{BAO-CMB Tension} \label{subsec: bao-cmb}

\begin{figure}
    \centering
    \includegraphics[width=0.75\linewidth]{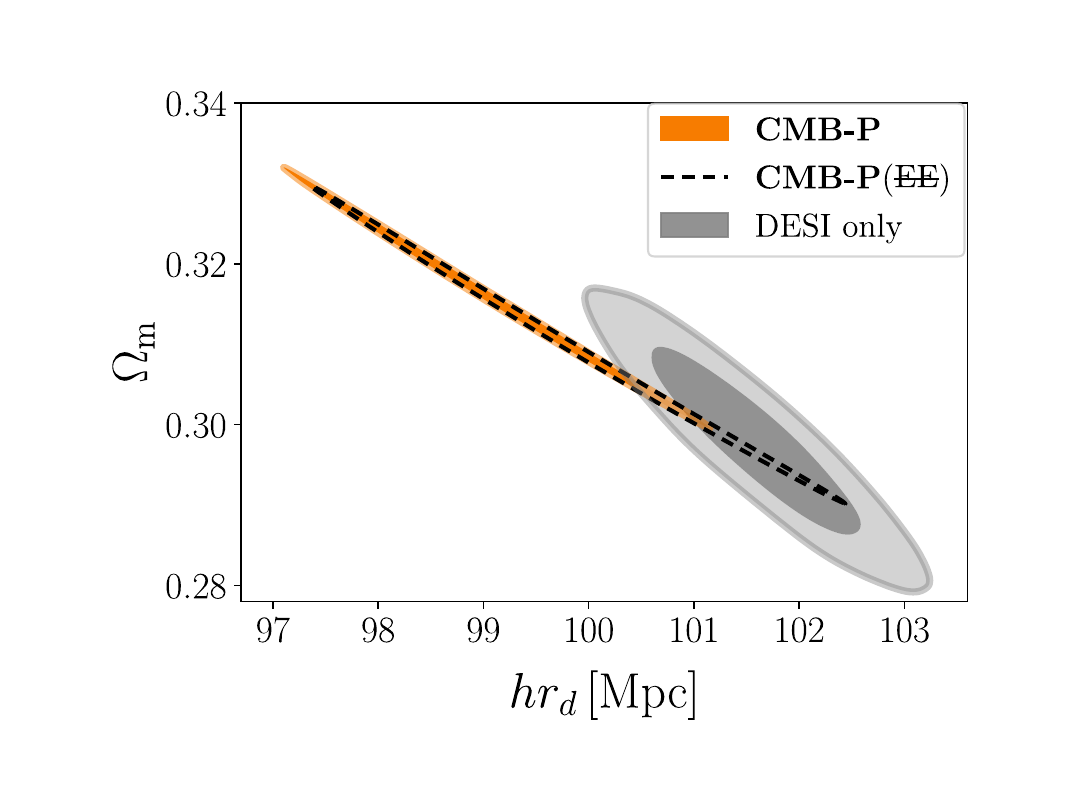}
    \caption{Two-dimensional posterior distribution of $\Omegam$ and $h r_d$ in $\Lambda$CDM. The 95\% C.L. constraints using  the \CMBP~and \CMBP\woee~datasets are depicted in the orange and black dashed contours, respectively, while the 68\% (95\%) C.L. contraints from DESI only are given by the dark (light) gray contours. }
    \label{fig:hrd_Omega_m}
\end{figure}

\begin{table}[]
    \centering
    \begin{tabular}{lcccccc}
\toprule
{\bf Pearson} & $\logA$ & $n_s$ & $h r_d \, [\mathrm{Mpc}]$ & $\omega_b$ & $\Omegam$ & $\taureio$ \\
\midrule
$\logA$ & 1.00 & 0.30 & 0.43 & 0.23 & -0.43 & 0.98 \\
$n_s$ & 0.30 & 1.00 & 0.50 & 0.23 & -0.50 & 0.33 \\
$h r_d \, [\mathrm{Mpc}]$ & 0.43 & 0.50 & 1.00 & 0.38 & -1.00 & \textbf{0.44} \\
$\omega_b$ & 0.23 & 0.23 & 0.38 & 1.00 & -0.43 & 0.22 \\
$\Omegam$ & -0.43 & -0.50 & -1.00 & -0.43 & 1.00 & \textbf{-0.44} \\
$\taureio$ & 0.98 & 0.33 & \textbf{0.44} & 0.22 & \textbf{-0.44} & 1.00 \\
\bottomrule
\end{tabular}

\begin{tabular}{lcccccc}
\toprule
{\bf Partial} & $\logA$ & $n_s$ & $h r_d \, [\mathrm{Mpc}]$ & $\omega_b$ & $\Omegam$ & $\taureio$ \\
\midrule
$\logA$ & --- & -0.13 & 0.10 & 0.15 & 0.10 & 0.98 \\
$n_s$ & -0.13 & --- & 0.02 & 0.03 & -0.00 & 0.16 \\
$h r_d \, [\mathrm{Mpc}]$ & 0.10 & 0.02 & --- & -0.71 & -1.00 & \textbf{-0.10} \\
$\omega_b$ & 0.15 & 0.03 & -0.71 & --- & -0.73 & -0.15 \\
$\Omegam$ & 0.10 & -0.00 & -1.00 & -0.73 & --- &\textbf{ -0.10} \\
$\taureio$ & 0.98 & 0.16 & \textbf{-0.10} & -0.15 & \textbf{-0.10} & --- \\
\bottomrule
\end{tabular}
    \caption{Pearson correlations (top) and partial correlations (bottom) between the parameters in the set $\{\logA,n_s,hr_d,\omega_b,\Omegam,\taureio\}$ of $\Lambda$CDM,  with the dataset \CMBP\woee+\bao+\pantheon. The correlations between $\taureio$ and $h r_d$ ($\Omegam$) are highlighted in bold.}
    \label{tab:hrdcorr}
\end{table}

\begin{figure}[h!]
    \centering
    \includegraphics[width=0.45\linewidth]{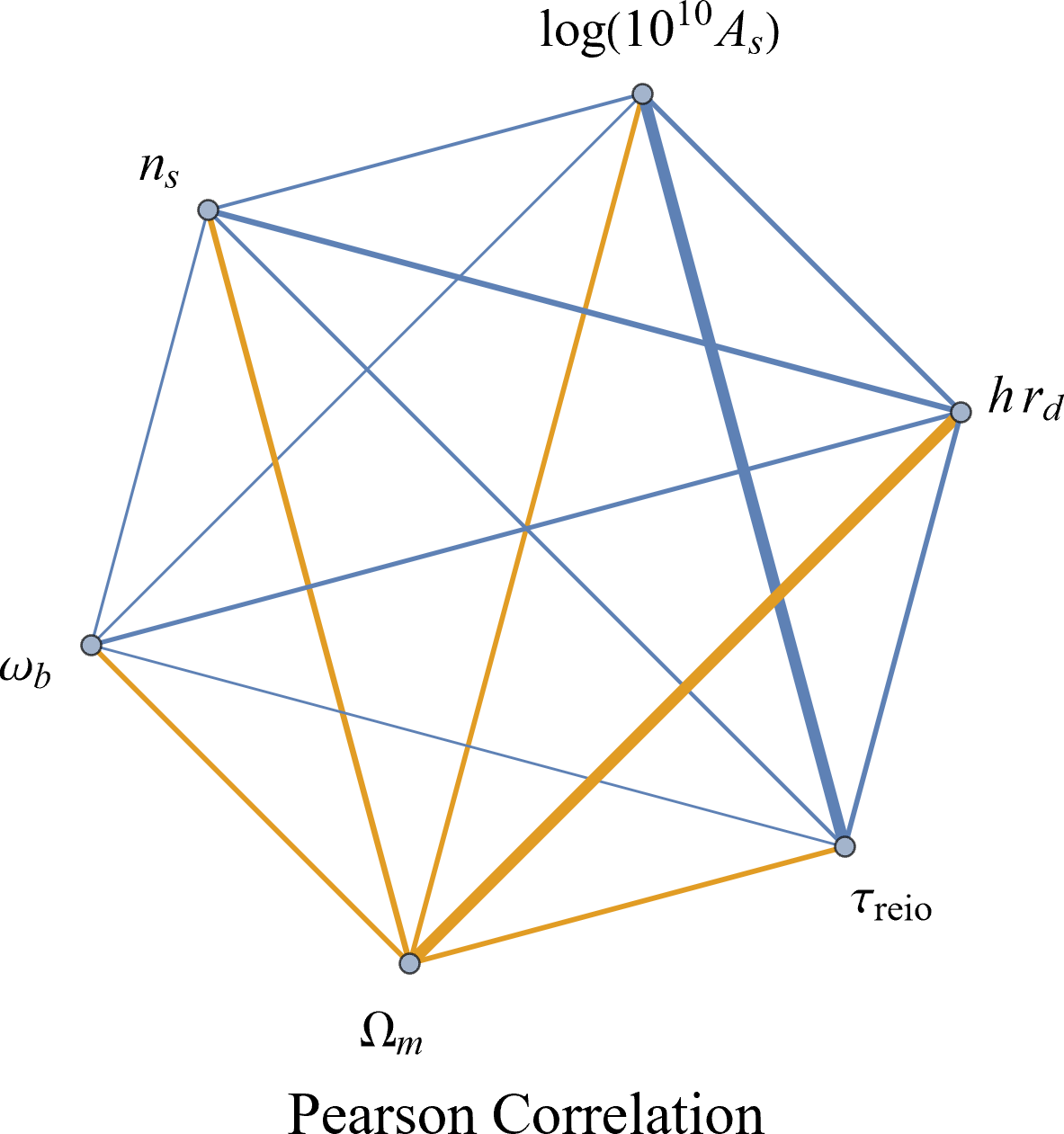}
    \includegraphics[width=0.45\linewidth]{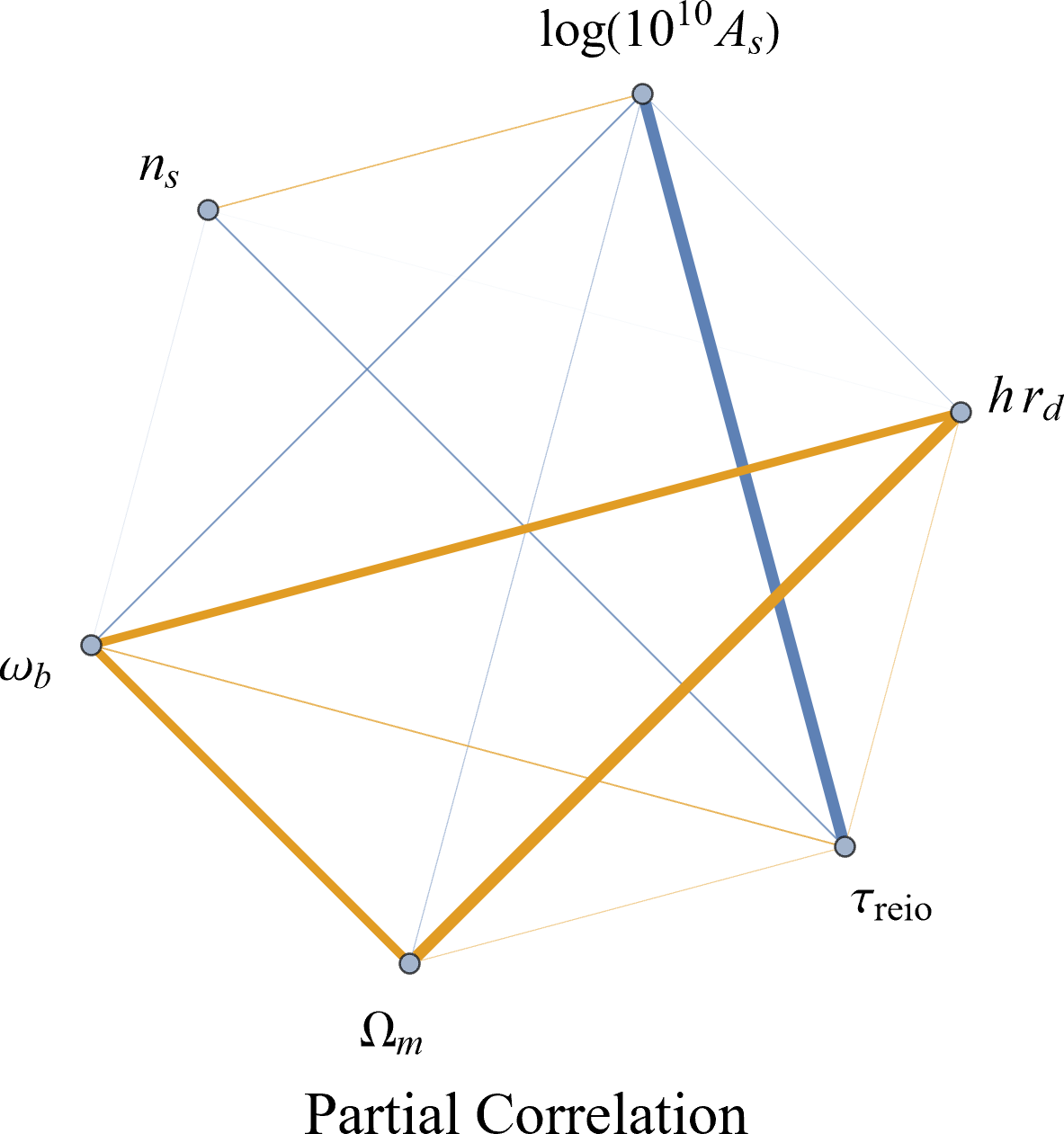}
    \caption{Pearson correlations (left) and partial correlations (right) between any two parameters of the set $\{\logA,n_s,hr_d,\omega_b,\Omegam,\taureio\}$ in $\Lambda$CDM, with the dataset \CMBP\woee+\bao+\pantheon. The plotting style is the same as in \cref{fig:corr_LCDM_1}.}
    \label{fig:corr_hrd}
\end{figure}

We now wish to comment on the mild tension between CMB datasets and the recent DESI BAO data. BAO is able to directly probe the expansion history of universe by constraining the total matter density today $\Omegam$, and the product $h r_d$, with $h = H_0/100$ and $r_d$ the physical size of the sound horizon at the baryon drag epoch. As pointed out in \cite{DESI:2025zgx,Ferreira:2025lrd}, there is a mild discrepancy between inferences from the CMB and from DESI BAO in this parameter space. The authors of \cite{Sailer:2025lxj} pointed out that this anomaly can also be correlated with the optical depth $\taureio$. In \cref{fig:hrd_Omega_m}, the tension between CMB and BAO data in these parameters could be visualized. One can see that the removal of the low-$\ell$ EE data and resulting increase in the inferred value of $\taureio$ lessens the tension, as the orange 2$\sigma$ contour shifts to the black dashed 2$\sigma$ contour, closer to the gray contours from DESI data alone.

Investigating the nature of the correlations between $\taureio$ and the parameters $\Omegam$ and $h r_d$, we once more compute the set of Pearson correlations and partial correlations, reported in \cref{tab:hrdcorr} (see \cref{app:PAS_corr} for correlations when using \CMBPAS). For the purposes of computing partial correlations, the set of parameters which are included is important as it determines which correlations are ``factored out" from each intrinsic relationship. For this reason, we choose a set of parameters which includes the parameters of interest $\{\Omegam,hr_d\}$ and is complementary to the set of free $\Lambda$CDM parameters. Thus, the parameter set is $\{\logA,n_s,hr_d,\omega_b,\Omegam,\taureio\}$. 

First, one can see clearly the Pearson correlations responsible for the relaxation of this tension. There is a positive correlation between $\taureio$ with $hr_d$ of 0.44, and a negative correlation between $\taureio$ and $\Omega_m$ of $-0.44$. These directions bring the inferences from the full dataset into closer agreement with the inferences made using only the DESI BAO data.
Examining, instead, the partial correlations, one sees once more that the relationship between this tension and $\taureio$ is not intrinsic to the parameters themselves, as the partial correlations are quite small. In the case of $hr_d$, the correlation with $\taureio$ is no longer in the correct direction switching from positive to negative. We can visualize these correlations in \cref{fig:corr_hrd}. The Pearson correlations of $\taureio$ and $\{\Omega_m,hr_d\}$ can be reconstructed with the partial correlations of other parameters. For example, the positive correlation of $hr_d$ and $\taureio$ receives a contribution from the path $\taureio\to\omega_b\to hr_d$, which contains two negative correlations and contributes positively.

Thus, similar to the Hubble tension in the previous section, we find a tension in data which is correlated with $\taureio$ and is mildly alleviated by larger values of $\taureio$. Again this relationship is found to be quite indirect, with the behavior of the other $\Lambda$CDM parameters being crucial to explain the overall correlation.

\subsection{Neutrino Mass Tension}\label{sec:nu_mass}

\begin{figure}
    \centering
    \includegraphics[width=0.75\linewidth]{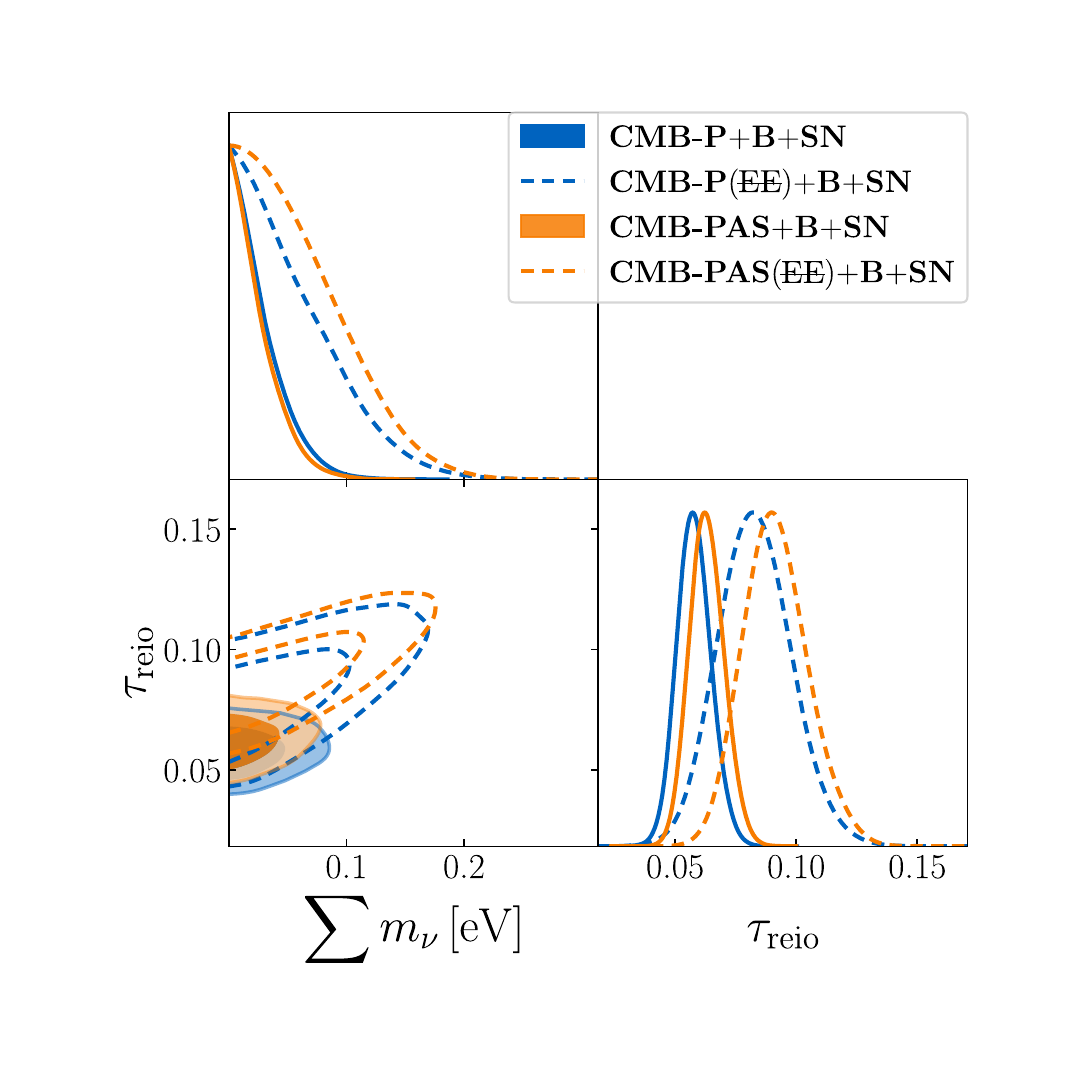}
    \caption{The one- and two-dimensional posterior distributions of $\taureio$ and $\sum m_\nu$ in a series of fits to four different combinations of datasets, assuming the $\Lambda$CDM+$\sum m_\nu$ model.  }
    \label{fig:mnu_tau}
\end{figure}

\begin{table}[]
    \centering
\begin{tabular}{lccccccc}
\toprule
 {\bf Pearson} & $\logA$ & $n_s$ & $H_0$ & $\omega_b$ & $\omegacdm$ & $\taureio$ & $\sum m_\nu$ \\
\midrule
$\logA$ & 1.00 & 0.47 & 0.02 & 0.34 & -0.63 & 0.99 & 0.60 \\
$n_s$ & 0.47 & 1.00 & 0.21 & 0.32 & -0.60 & 0.48 & 0.41 \\
$H_0$ & 0.02 & 0.21 & 1.00 & 0.40 & -0.32 & \textbf{0.03} & \textbf{-0.45} \\
$\omega_b$ & 0.34 & 0.32 & 0.40 & 1.00 & -0.36 & 0.32 & 0.27 \\
$\omegacdm$ & -0.63 & -0.60 & -0.32 & -0.36 & 1.00 & -0.63 & -0.63 \\
$\taureio$ & 0.99 & 0.48 & \textbf{0.03} & 0.32 & -0.63 & 1.00 & \textbf{0.58} \\
$\sum m_\nu$ & 0.60 & 0.41 & \textbf{-0.45} & 0.27 & -0.63 & \textbf{0.58} & 1.00 \\
\bottomrule
\end{tabular}

\begin{tabular}{lccccccc}
\toprule
 {\bf Partial} & $\logA$ & $n_s$ & $H_0$ & $\omega_b$ & $\omegacdm$ & $\taureio$ & $\sum m_\nu$ \\
\midrule
$\logA$ & --- & -0.13 & 0.09 & 0.01 & 0.09 & 0.98 & 0.15 \\
$n_s$ & -0.13 & --- & 0.08 & 0.01 & -0.07 & 0.16 & 0.09 \\
$H_0$ & 0.09 & 0.08 & --- & 0.80 & -0.92 & \textbf{-0.09} & \textbf{-0.96} \\
$\omega_b$ & 0.01 & 0.01 & 0.80 & --- & 0.70 & -0.00 & 0.76 \\
$\omegacdm$ & 0.09 & -0.07 & -0.92 & 0.70 & --- & -0.11 & -0.92 \\
$\taureio$ & 0.98 & 0.16 & \textbf{-0.09} & -0.00 & -0.11 & --- & \textbf{-0.13} \\
$\sum m_\nu$ & 0.15 & 0.09 & \textbf{-0.96} & 0.76 & -0.92 & \textbf{-0.13} & --- \\
\bottomrule
\end{tabular}
    \caption{Pearson correlations (top) and partial correlations (bottom) between parameters of $\Lambda$CDM$+\sum m_\nu$, with the dataset \CMBP\woee+\bao+\pantheon. The correlations between $\taureio$ and $\sum m_\nu$ ($H_0$) are highlighted in bold. }
    \label{tab:mnucorr}
\end{table}

\begin{figure}[h!]
    \centering
    \includegraphics[width=0.45\linewidth]{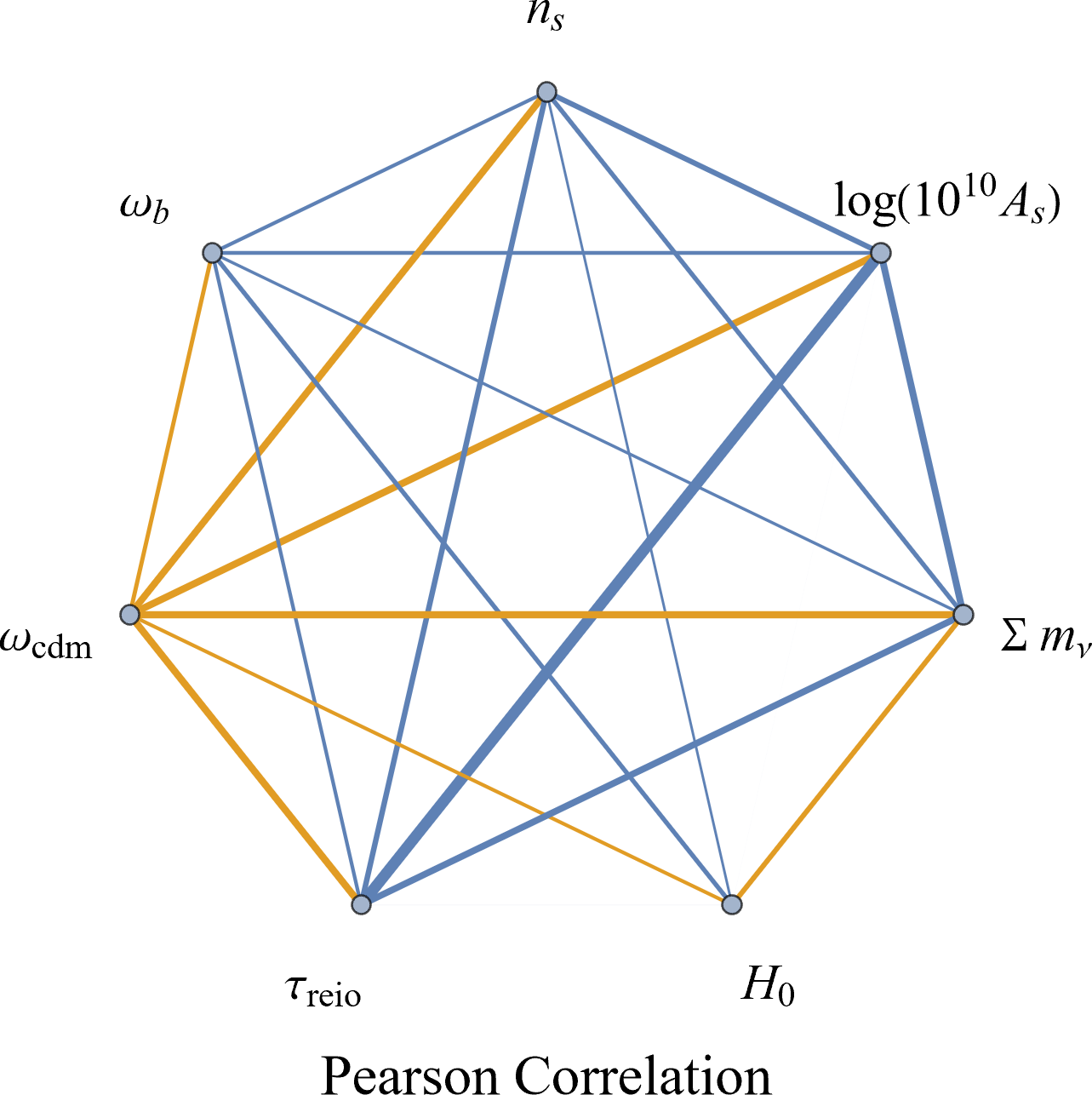}
    \hspace{4mm}
    \includegraphics[width=0.45\linewidth]{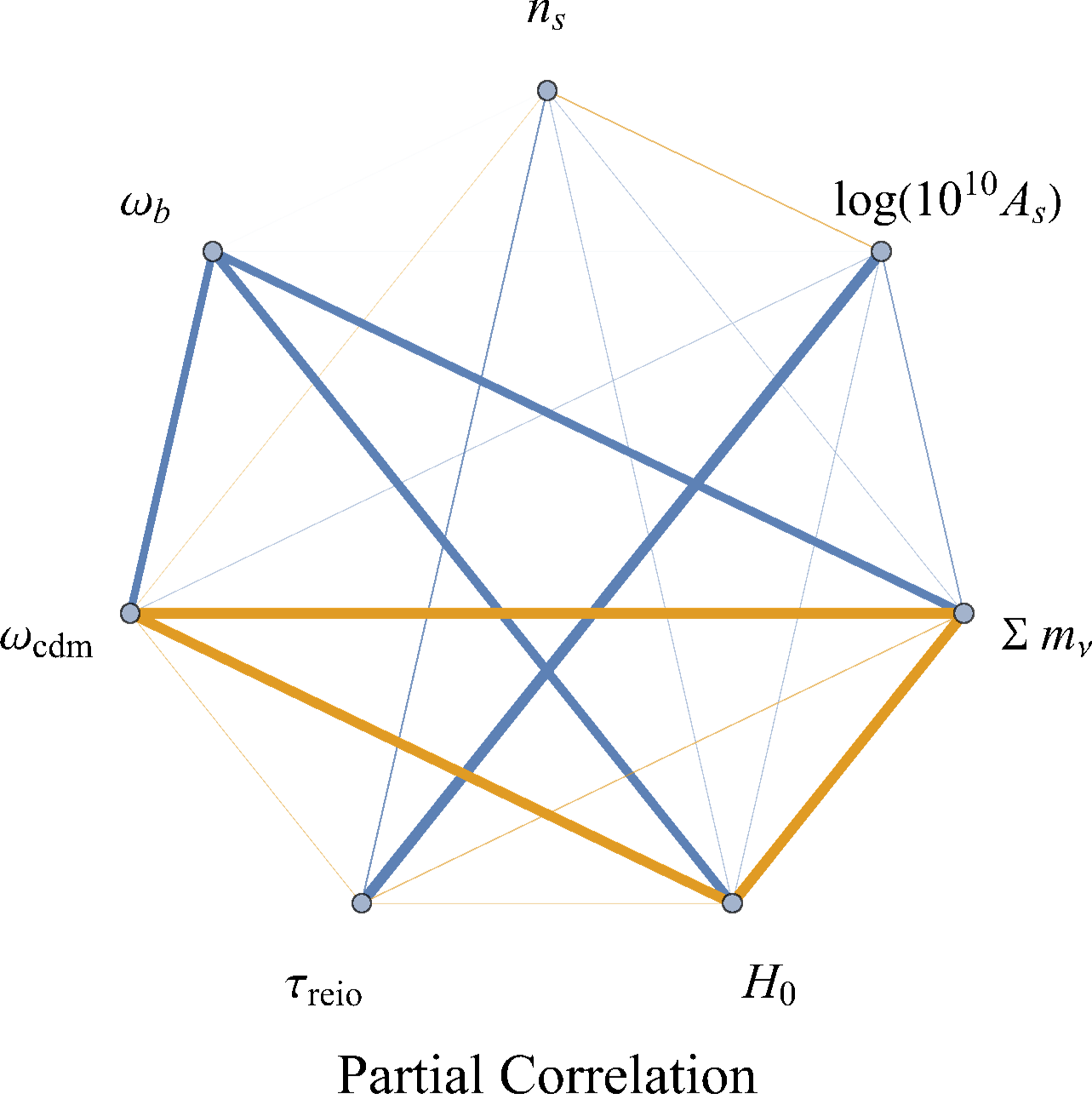}
    \caption{Pearson correlations (left) and partial correlations (right) between parameters of $\Lambda$CDM plus $\sum m_\nu$, with the dataset \CMBP\woee+\bao+\pantheon. The plotting style is the same as in \cref{fig:corr_LCDM_1}. 
    }
    \label{fig:corr_Mnu_1}
\end{figure}

Another anomalous finding in recent cosmological data, the tight constraints on the sum of neutrino masses $\sum m_\nu$, is also correlated with $\taureio$. In the context of adding the neutrino mass sum (with a prior $\sum m_\nu >0$) as one additional parameter to $\Lambda$CDM, which we will refer to as $\Lambda$CDM+$\sum m_\nu$, upper bounds on $\sum m_\nu$ involving the newest BAO data from DESI DR2 \cite{DESI:2025zgx} are in tension with lower bounds from neutrino oscillation experiments, about 0.06~eV for the normal hierarchy case and 0.1~eV for the inverted hierarchy case~\cite{ParticleDataGroup:2024cfk}. 

As pointed out in \cite{Jhaveri:2025neg,Sailer:2025lxj}, an increase in $\taureio$ can alleviate this tension slightly. This could be seen from \cref{fig:mnu_tau}: when the low-$\ell$ EE data is removed, $\taureio$ increases together with a relaxation of the upper bound on $\sum m_\nu$. The bound relaxes to $\sum m_\nu < 0.13$ eV for \CMBP\woee+\bao+\pantheon, easily accommodating lower bounds from neutrino oscillations experiments. Of particular interest is that, with the addition of ACT and SPT CMB observations, the constraint on $\sum m_\nu$ inferred from the \CMBPAS\woee+\bao+\pantheon~dataset relaxes further to $\sum m_\nu < 0.14$ eV, and the posterior starts to form a peak. Although this peak does not constitute a statistically significant lower bound, it suggests that a larger inferred $\taureio$ could indeed lead to detection of a nonzero neutrino mass sum with future updates to CMB observations. In addition, it has been shown in the literature \cite{Allali:2024aiv,Loverde:2024nfi} that the constraints on neutrino mass are sensitive to the choice of supernovae dataset. In \cref{app:nu_mass_DES}, we present similar results when using the \DES~dataset, showing a posterior with a slightly more prominent peak away from zero neutrino mass sum.

Similar to before, we assess the correlations of $\taureio$ by computing the Pearson and partial correlations of the parameters in the $\Lambda$CDM$+\sum m_\nu$ model, shown in \cref{tab:mnucorr} (see \cref{app:PAS_corr} for correlations when using \CMBPAS). As expected according to its definition, the partial correlation between the six $\Lambda$CDM parameters remain largely intact in \cref{tab:LCDMcorr} and \cref{tab:mnucorr}. In other words, the inclusion of $\Sigma m_\nu$ in the fit has been properly factored out when obtaining partial correlations. We can also see that while the Pearson correlation between $\taureio$ and $\sum m_\nu$ is relatively strong at 0.58, it becomes small and negative when considering the partial correlation. Thus, this alleviation of the neutrino-mass tension by a larger $\taureio$ is an indirect effect relating to other parameters of the model. Examining \cref{fig:corr_Mnu_1}, one can attempt to construct the parameter relationships that are responsible for the eventual relationship between $\taureio$ and $\sum m_\nu$. For instance, the path $\taureio\to\omegacdm\to\sum m_\nu$ contributes positively.

Yet another interesting observation can be made when examining the correlations of $H_0$ in this model. Note that $H_0$ and $\sum m_\nu$ are tightly negatively correlated, both in the Pearson correlation (-0.45) and especially in the partial correlation (-0.96). Meanwhile, as discussed in \cref{sec:H0tension}, $H_0$ and $\taureio$ are positively correlated in the $\Lambda$CDM model. Interestingly, this positive correlation is functionally erased by the strong anti-correlation between $H_0$ and $\sum m_\nu$, such that in the $\Lambda$CDM$+\sum m_\nu$ model, $\taureio$ does not correlate with $H_0$ appreciably. This seems to suggest that the ultimate correlations of $\taureio$ with either $H_0$ or $\sum m_\nu$ rely on similar parameters, while the correlation between $\taureio$ with $\sum m_\nu$ is somewhat stronger such that the $\taureio$-$H_0$ correlation doesn't survive when $\sum m_\nu$ is a free parameter.

\subsection{Nature of Dark Energy}\label{sec:w0wa}

\begin{figure}
    \centering
    \includegraphics[width=0.75\linewidth]{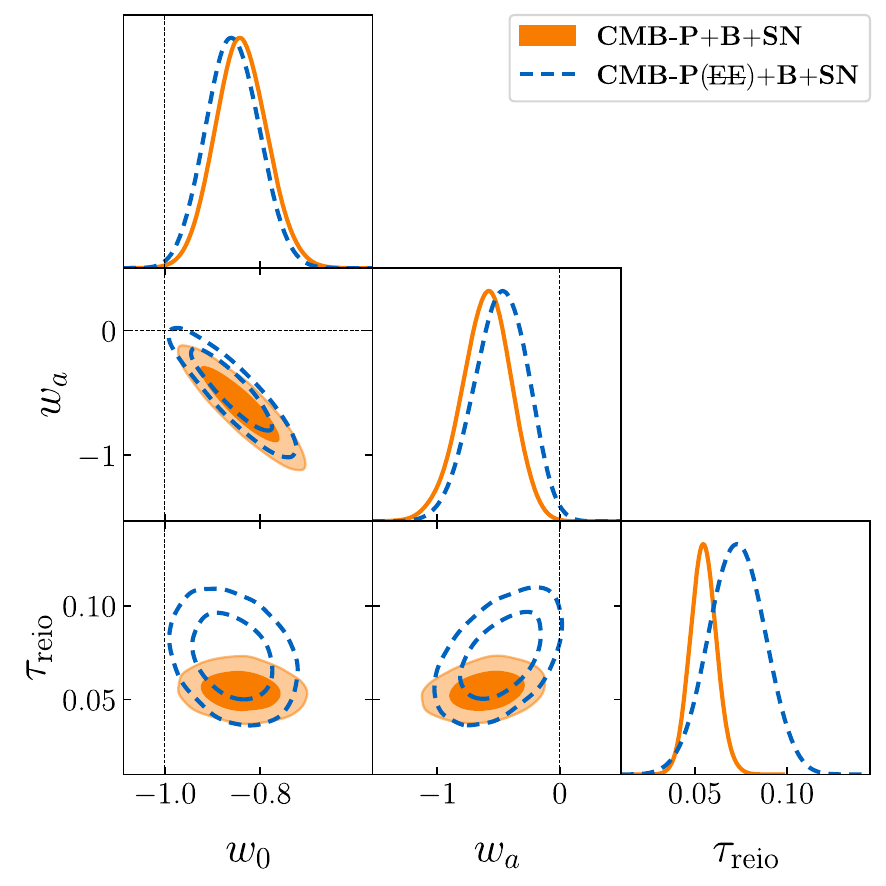}
    \caption{The one- and two-dimensional posterior distributions of $\taureio$, $w_0$, and $w_a$ in a series of fits to two different combinations of datasets, assuming the $w_0w_a$CDM model. }
    \label{fig:w0wa_tau}
\end{figure}

\begin{table}[]
    \centering
\begin{tabular}{lcccccccc}
\toprule
{\bf Pearson} & $\logA$ & $n_s$ & $H_0$ & $\omega_b$ & $\omegacdm$ & $\taureio$ & $w_0$ & $w_a$ \\
\midrule
$\logA$ & 1.00 & 0.52 & -0.04 & 0.45 & -0.66 & 0.99 & -0.29 & 0.49 \\
$n_s$ & 0.52 & 1.00 & 0.00 & 0.43 & -0.66 & 0.53 & -0.27 & 0.45 \\
$H_0$ & -0.04 & 0.00 & 1.00 & 0.08 & 0.05 & -0.04 & -0.54 & 0.16 \\
$\omega_b$ & 0.45 & 0.43 & 0.08 & 1.00 & -0.51 & 0.44 & -0.19 & 0.36 \\
$\omegacdm$ & -0.66 & -0.66 & 0.05 & -0.51 & 1.00 & -0.66 & 0.34 & -0.62 \\
$\taureio$ & 0.99 & 0.53 & -0.04 & 0.44 & -0.66 & 1.00 & \textbf{-0.28} & \textbf{0.48} \\
$w_0$ & -0.29 & -0.27 & -0.54 & -0.19 & 0.34 & \textbf{-0.28} & 1.00 & -0.89 \\
$w_a$ & 0.49 & 0.45 & 0.16 & 0.36 & -0.62 & \textbf{0.48} & -0.89 & 1.00 \\
\bottomrule
\end{tabular}

\begin{tabular}{lcccccccc}
\toprule
{\bf Partial} & $\logA$ & $n_s$ & $H_0$ & $\omega_b$ & $\omegacdm$ & $\taureio$ & $w_0$ & $w_a$ \\
\midrule
$\logA$ & --- & -0.10 & 0.06 & 0.04 & 0.07 & 0.98 & 0.07 & 0.08 \\
$n_s$ & -0.10 & --- & 0.07 & 0.04 & -0.10 & 0.13 & 0.06 & 0.07 \\
$H_0$ & 0.06 & 0.07 & --- & 0.76 & -0.89 & -0.06 & -0.99 & -0.98 \\
$\omega_b$ & 0.04 & 0.04 & 0.76 & --- & 0.64 & -0.02 & 0.76 & 0.76 \\
$\omegacdm$ & 0.07 & -0.10 & -0.89 & 0.64 & --- & -0.09 & -0.91 & -0.92 \\
$\taureio$ & 0.98 & 0.13 & -0.06 & -0.02 & -0.09 & --- & \textbf{-0.06} & \textbf{-0.07} \\
$w_0$ & 0.07 & 0.06 & -0.99 & 0.76 & -0.91 & \textbf{-0.06} & --- & -1.00 \\
$w_a$ & 0.08 & 0.07 & -0.98 & 0.76 & -0.92 & \textbf{-0.07} & -1.00 & --- \\
\bottomrule
\end{tabular}
    \caption{Pearson correlations (top) and partial correlations (bottom) between parameters in $w_0w_a$CDM, with the dataset \CMBP\woee+\bao+\pantheon. The correlations between $\taureio$ and $w_0$ ($w_a$) are highlighted in bold.}
    \label{tab:w0wacorr}
\end{table}

\begin{figure}[h!]
    \centering
    \includegraphics[width=0.45\linewidth]{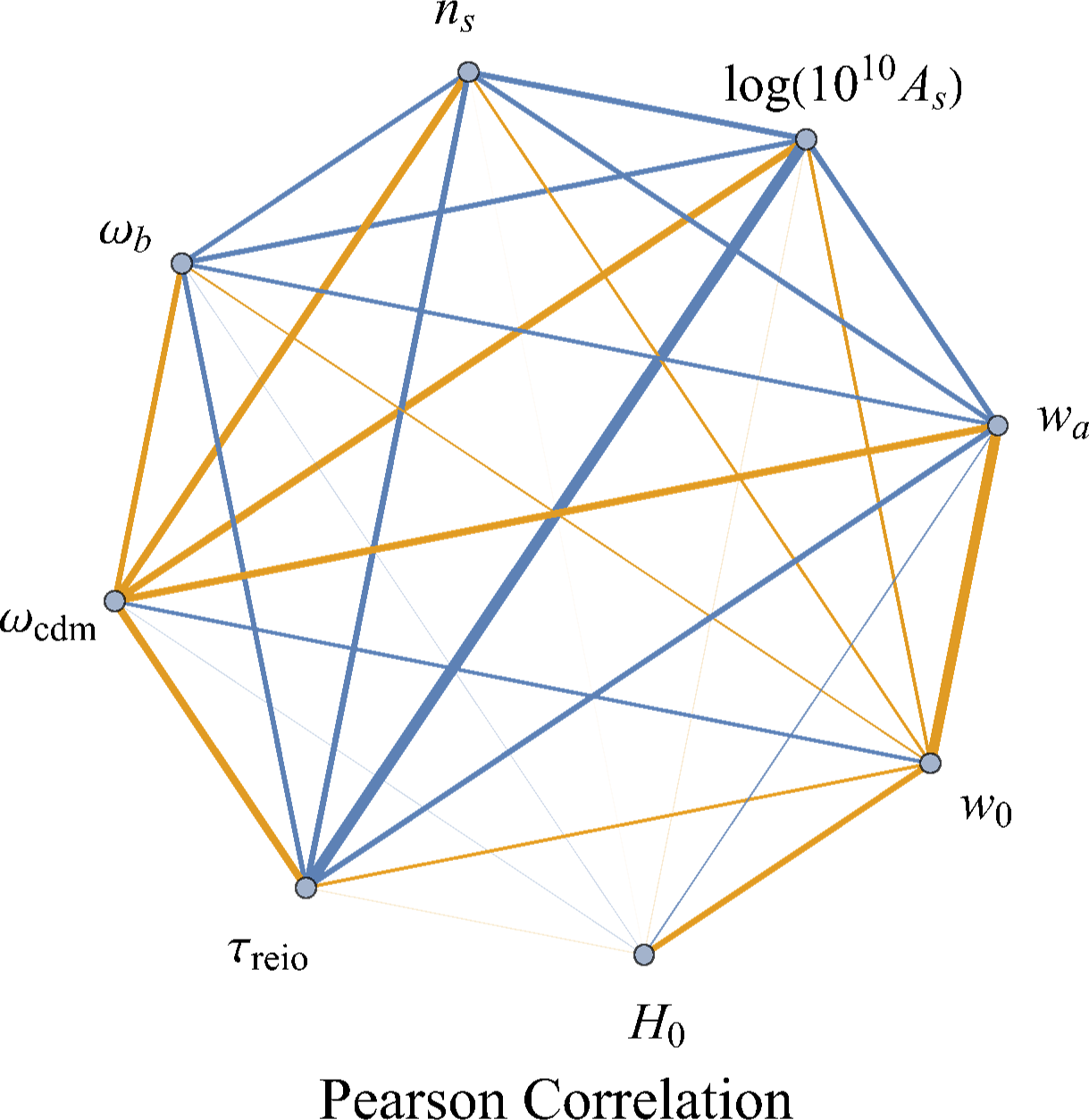}
    \includegraphics[width=0.45\linewidth]{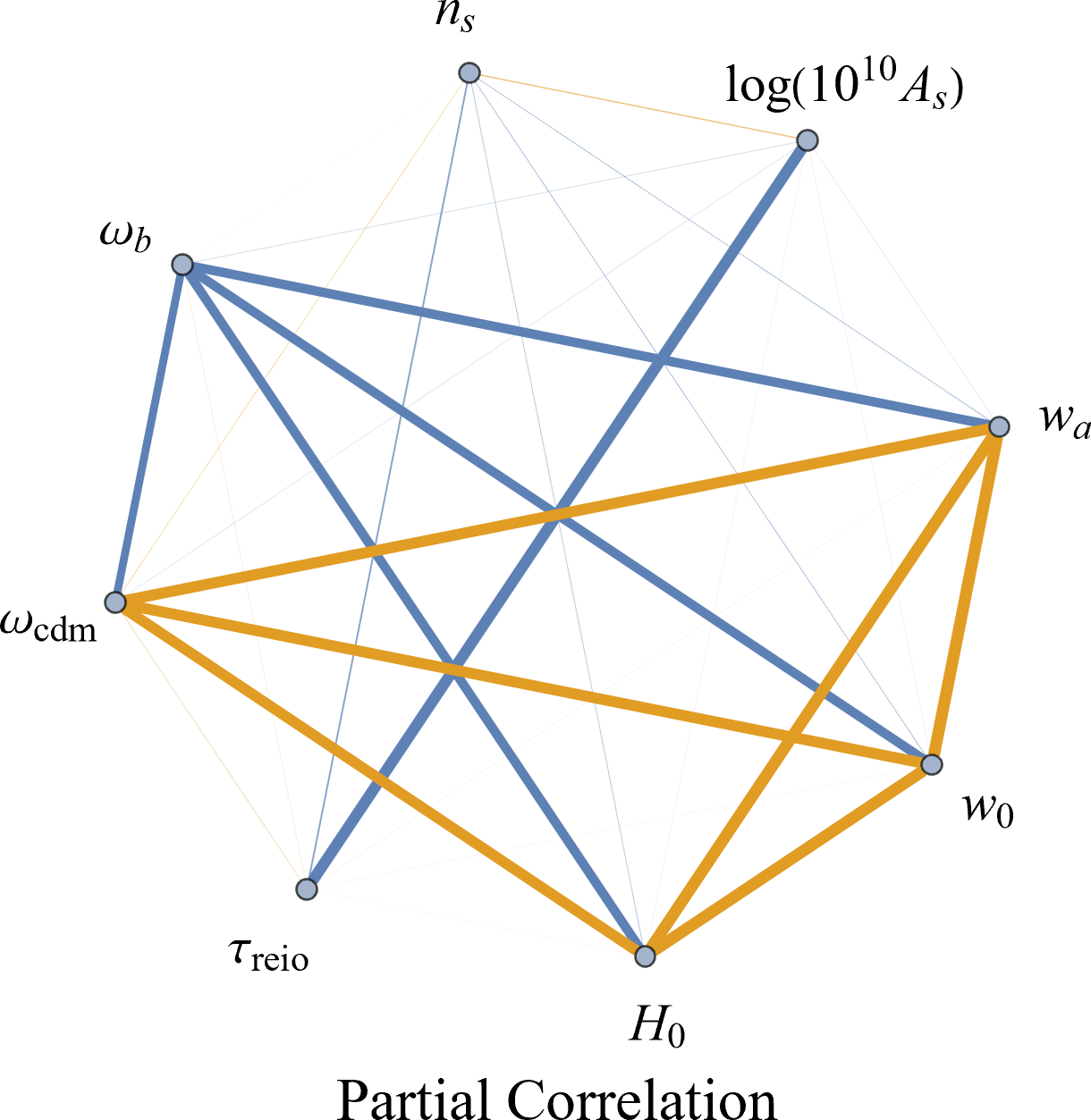}
    \caption{Pearson correlations (left) and partial correlations (right) between parameters of $w_0w_a$CDM, with the dataset \CMBP\woee+\bao+\pantheon. The plotting style is the same as in \cref{fig:corr_LCDM_1}.}
    \label{fig:corr_wowa_1}
\end{figure}

In addition to the tensions/anomalies discussed so far, recent analyses involving the BAO measurements from DESI in combination with several datasets of supernovae have challenged the cosmological constant paradigm for DE. Rather than a constant vacuum-energy-like component $\Lambda$, there is mild-to-moderate evidence in the data for a dynamical nature of DE. One widely-studied parameterization, the CPL parameterization, assumes the equation of state of dark energy is $w(a)=w_0+w_a(1-a)$, where $w_0$, $w_a$ are constants, $a$ is the scale factor, and $(w_0, w_a) = (-1,0)$ corresponds to a cosmological constant DE. Depending on the choice of supernovae dataset, DESI BAO data leads to an approximately 2-4$\sigma$ evidence for deviations from a cosmological constant, or equivalently $(w_0, w_a)$ departing from $(-1, 0)$ \cite{DESI:2025zgx,DESI:2025wyn}.

The authors of \cite{Sailer:2025lxj} pointed out that this behavior may also be mildly correlated with the optical depth $\taureio$. Focusing on the CPL parameterization, which we will refer to as the $w_0w_a$CDM model, we present posteriors in \cref{fig:w0wa_tau} demonstrating that the deviation from $\Lambda$CDM is marginally reduced with a larger $\taureio$ (by switching from the \CMBP+\bao+\pantheon~to \CMBP\woee+\bao+\pantheon~dataset). We present the correlations in \cref{tab:w0wacorr} and \cref{fig:corr_wowa_1}. One can remark yet again that the mild-to-moderate Pearson correlations of $w_0$ and $w_a$ with $\taureio$ are almost absent in the partial correlations, suggesting an indirect nature to this relationship. More specifically, the partial correlation between $\taureio$ and $w_0$ as well as $w_a$ are both negative and small. Yet through all possible combinations of paths in the partial correlation diagram, the Pearson correlation between $\taureio$ and $w_a$ becomes enhanced and positive while the $\taureio-w_0$ correlation remains negative but also enhanced in its absolute value. We find that replacing the supernovae dataset \pantheon~by either \DES~or \union~does not alter the correlations significantly, even if it changes the significance of detecting dynamical DE (see \cref{app:DES_union} for more details).

Another interesting feature is the similarity between $w_0$ and $w_a$, which is most obvious in the partial correlations. In particular, the partial correlations between $w_0$/$w_a$ and other cosmological parameters are almost identical to each other. This suggests that in the data, the perturbations of $w_0$ and $w_a$ around the best fit point are highly indistinguishable, suggesting that the data are most sensitive to a linear combination of $w_0$ and $w_a$, rather than each parameter individually. This point can also be seen in the extreme partial correlation of $\sim -0.997$ between $w_0$ and $w_a$.
Turning to the physical meaning, we can remark that the equation of state of DE is indeed a linear combination of $w_0$ and $w_a$ for a given redshift. Thus, the tightness of their correlation suggests that the inferred results are most sensitive to a narrow range of redshifts, corresponding to a particular linear combination. However, this twin-behavior between $w_0$ and $w_a$ is less obvious in the Pearson correlations, which is also due to the strong negative partial correlation between $w_0$ and $w_a$. For any cosmological parameter $X$, its Pearson correlation with $w_0$ receives contributions from the partial correlation $w_0\to X$ and the path through $w_a$, $i.e.$, $w_0\to w_a \to X$. The two major contributors will almost cancel each other due to the strong negative partial correlation between $w_0$ and $w_a$, such that the Pearson correlation of $X$ with either $w_0$ or $w_a$ depends instead on small differences and more convoluted paths. Thus, the Pearson correlations end up quite different. The non-trivial relationship between $w_0$ and $w_a$ can be derived starting from the Pearson correlations, but it is far from apparent from this point of view. Such a hidden structure is, again, revealed in the dual description of partial correlations.

\section{Effects of Datasets} 
\label{sec: diff data}

From the discussion above, we demonstrate that cosmological parameters are correlated in a nested way. Either the Pearson or partial correlations appear as networks between cosmological parameters. Loop structures in, $e.g.$, \cref{fig:corr_LCDM_1}, prevent us from easily identifying the causes and effects among cosmological parameters. In contrast, it is unambiguous to claim that the inclusion or exclusion of a certain dataset causes the cosmological parameters' distributions to change.

\begin{figure}[h!]
    \centering
    \includegraphics[width=0.45\linewidth]{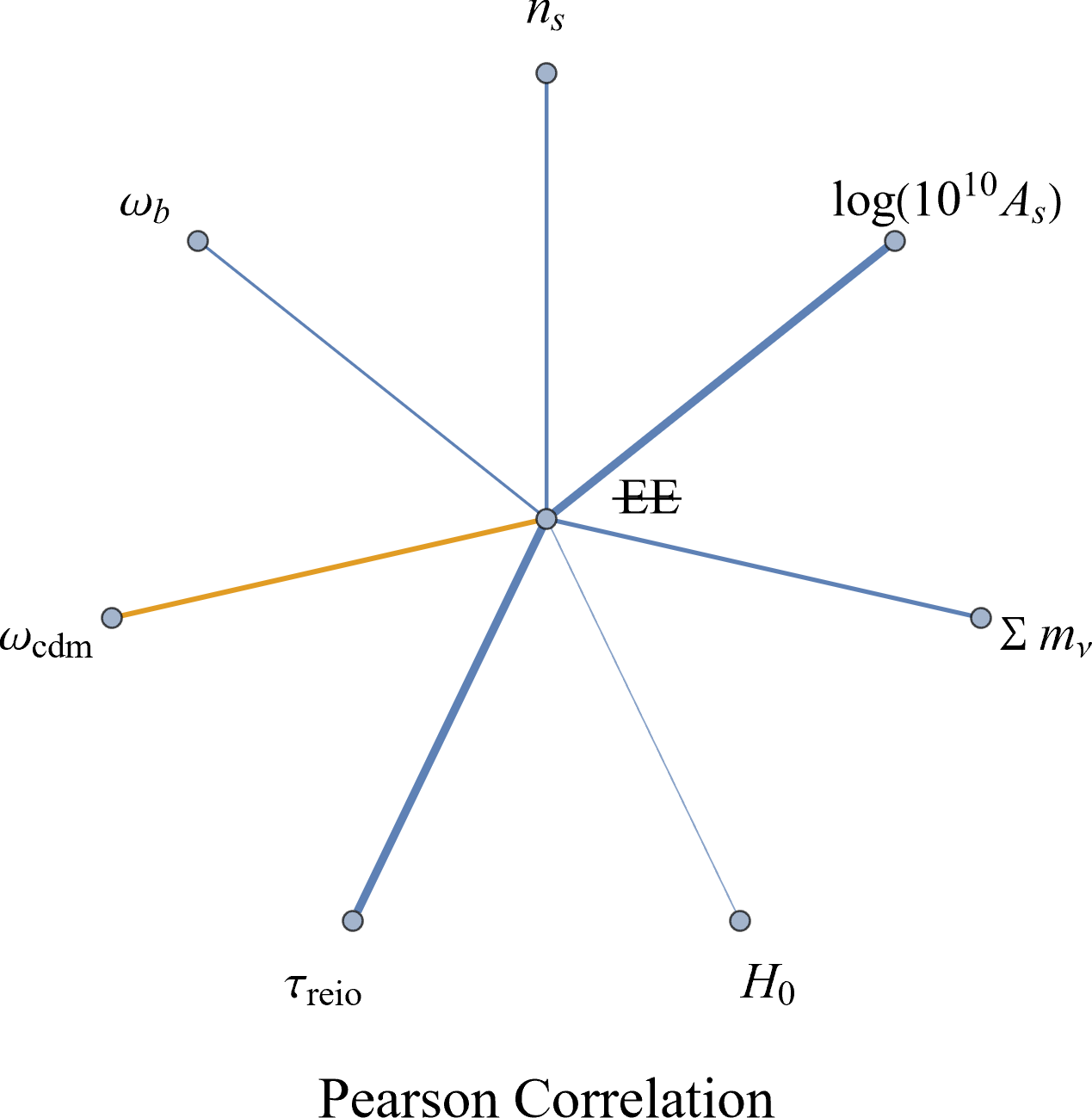}
    \hspace{4mm}
    \includegraphics[width=0.45\linewidth]{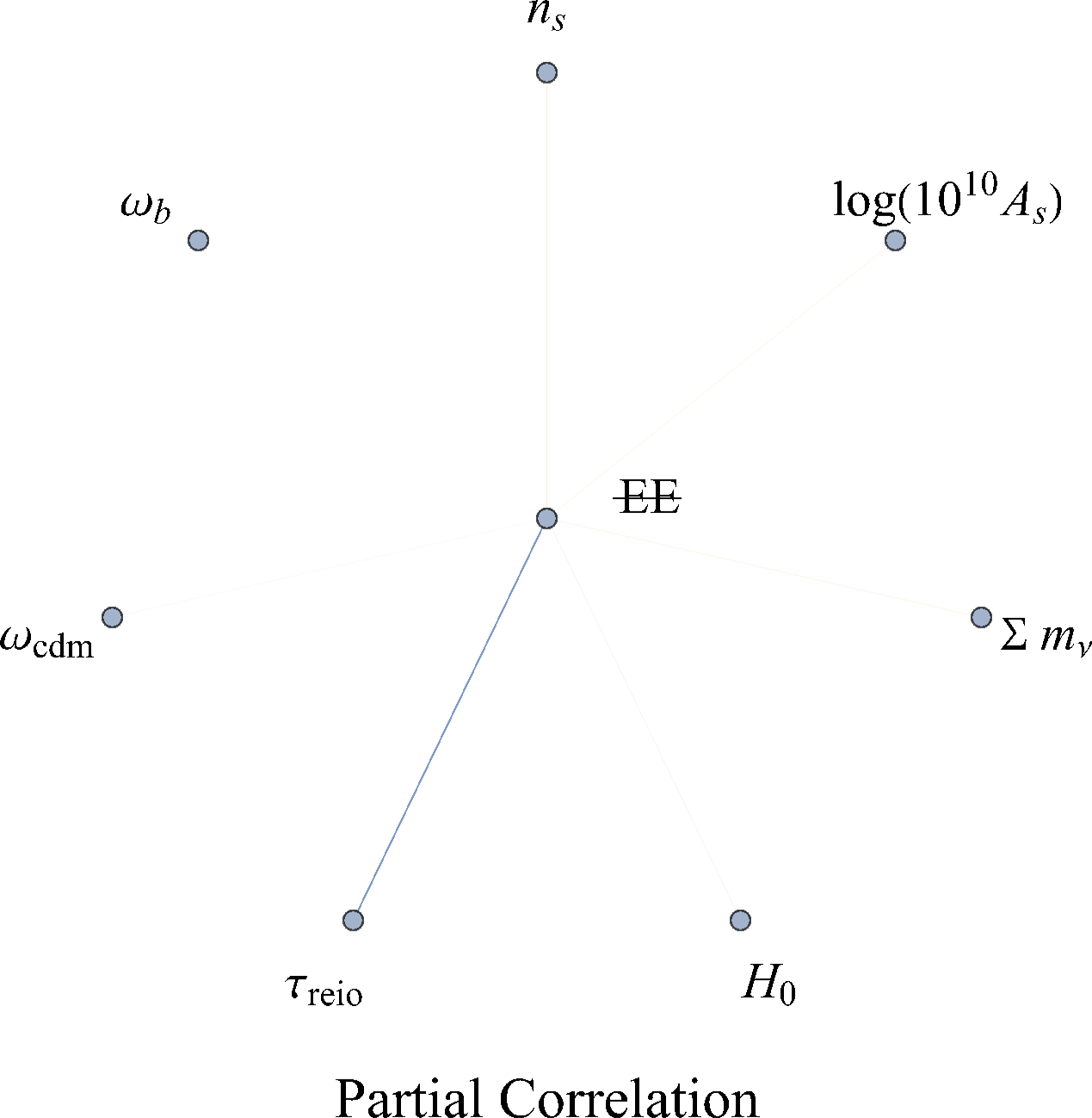}
    \caption{Correlation diagrams highlighting parameter correlations with the exclusion of the low-$\ell$  EE data of Planck \woee. The left and right panels are the Pearson and partial correlations, respectively. Blue lines indicate positive correlations, orange lines indicate negative, and the line thickness is broadcast to a linear scale of the correlation strength. See the main text for more details.}
    \label{fig:corr_EECas_1}
\end{figure}

In this section, we want to comment on the effects of certain datasets on inferring the cosmological parameters and their correlations. Based on the discussions developed in \cref{sec:results_all}, we analyze the effects of a given dataset by combining the MC samples obtained with and without that dataset, while properly labeling each point in the sample as either including the dataset of interest, or not.
By treating the dataset label as a binary parameter 
of the MC sample,
one can calculate the Pearson and partial correlation between this binary parameter and other cosmological parameters.\footnote{The method is inspired by the field of causal inference~\cite{Pearl2016CausalII}, where the inclusion (exclusion) of a new dataset could be considered as a ``treatment" on the cosmology, and the effect of such a treatment can be inferred from the distribution differences of cosmological parameters.}

\begin{figure}[]
    \centering
    \includegraphics[width=0.45\linewidth]{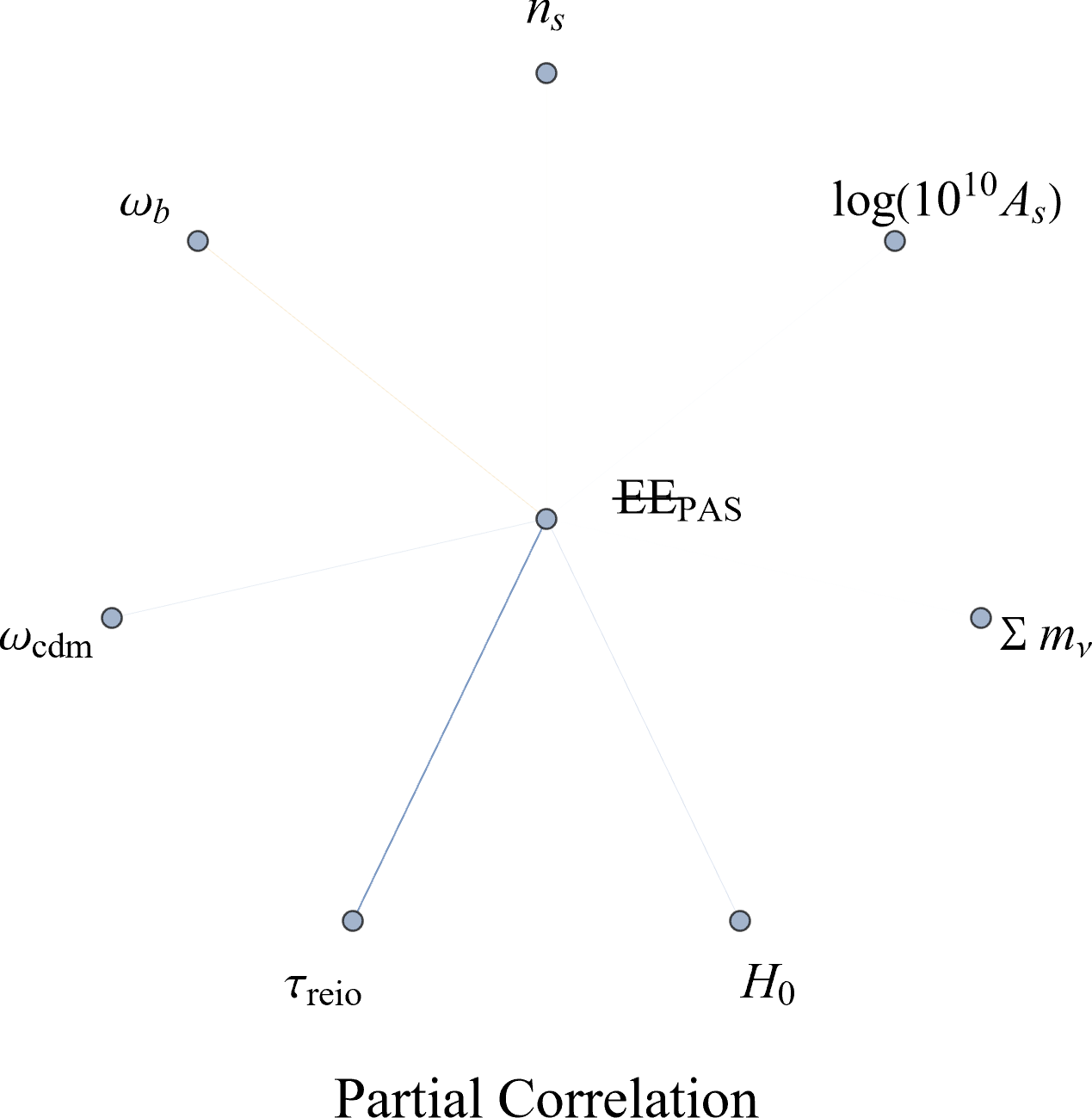}
    \caption{Partial correlation diagram highlighting parameter correlations with the exclusion of the low-$\ell$  EE data of Planck \woee, within the combination \CMBPAS+\bao+\pantheon. An appreciable correlation is only observed for $\taureio$, as is the case with \CMBP, \CMBPA, and \CMBPS. The plotting style is the same as \cref{fig:corr_EECas_1}.}
    \label{fig:corr_EECas_PAS_1}
\end{figure}

We first show the impact of excluding the low-$\ell$ EE data from Planck to all six cosmological parameters of $\Lambda$CDM, plus the total neutrino mass. A joint MC sample is constructed by taking an equal weight combination of the \CMBP+\bao+\pantheon~and the \CMBP\woee+\bao+\pantheon~MC samples. All \woee~samples are labeled with a binary parameter with value one, while the others are labeled with value zero. The Pearson and partial correlations of this binary parameter and other cosmological parameters are calculated, and the results are presented in \cref{fig:corr_EECas_1}. In both panels, only the correlation between the binary \woee~parameter and other cosmological parameters is shown, with the former node placed in the middle of each panel for better display. The Pearson correlation directly describes the effect of excluding the low-$\ell$ EE data. For instance, if using the \CMBP\woee~dataset increases a parameter such as $\taureio$, the Pearson correlation between $\taureio$ and the binary label parameter is then positive; if the result is to decrease the parameter, the correlation is negative. The left panel of 
\cref{fig:corr_EECas_1} shows that excluding the low-$\ell$ EE data affects all cosmological parameters, with most of them trending towards larger values, except for $\omegacdm$. The point has been thoroughly discussed in~\cite{Allali:2025wwi} and~\cite{Giare:2023ejv}. On the contrary, in the dual view of partial correlations, the effect of \woee~dataset is more straightforward to be understood. As first pointed out by~\cite{Allali:2025wwi} and also in \cref{sec:results_all}, excluding the low-$\ell$ EE data does not affect most cosmological parameters other than $\taureio$ directly. Instead, the effects are seen as the Pearson correlations between the \woee~and cosmological parameters, including $A_s$, largely stem from increasing $\taureio$, as demonstrated by the partial correlations in the right panel of 
\cref{fig:corr_EECas_1}. Other than $\taureio$ with a partial correlation of 0.16, the binary label parameter of \woee~develops no significant partial correlation with other cosmological parameters, with the second most significant one being with $n_s$ at less than 0.04. The relation strengthens the argument that the low-$\ell$ EE part of the Planck dataset breaks the degeneracy between $A_s$ and $\taureio$, constraining the latter to a lower value and then affecting other cosmological parameters. To further validate the points above, we also show the effect of excluding low-$\ell$ EE data when using the $\CMBPAS$ datasets in the same way. The partial correlations are shown in \cref{fig:corr_EECas_PAS_1}. In this case, the direct effect of removing the low-$\ell$ EE data is highly consistent with the $\CMBP$ case, with the leading effect of raising $\taureio$.  We have checked that the same consistency is seen when using \CMBPA~or \CMBPS~as well.

\begin{figure}[h!]
    \centering
    \includegraphics[width=0.45\linewidth]{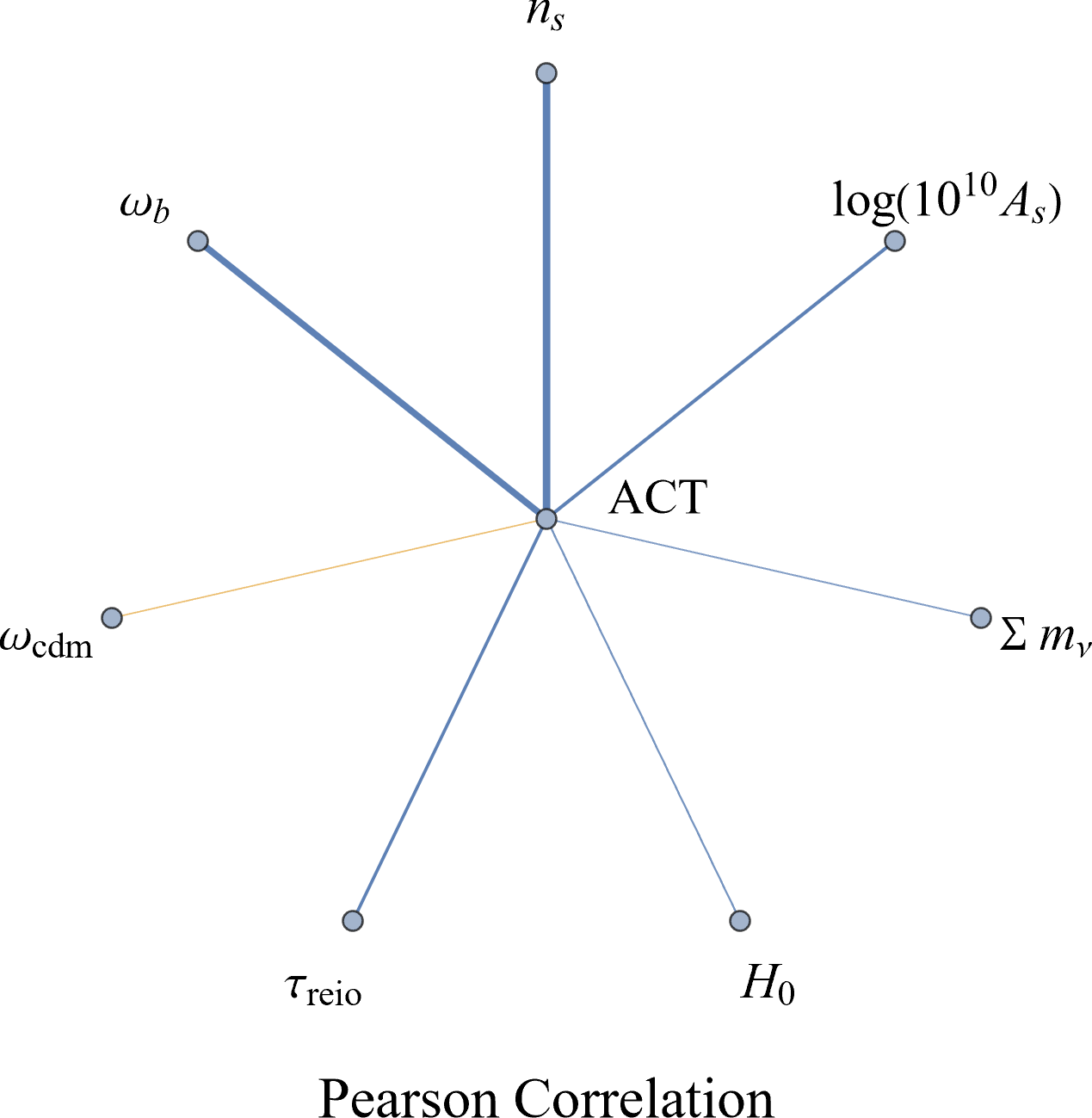}
    \hspace{4mm}
    \includegraphics[width=0.45\linewidth]{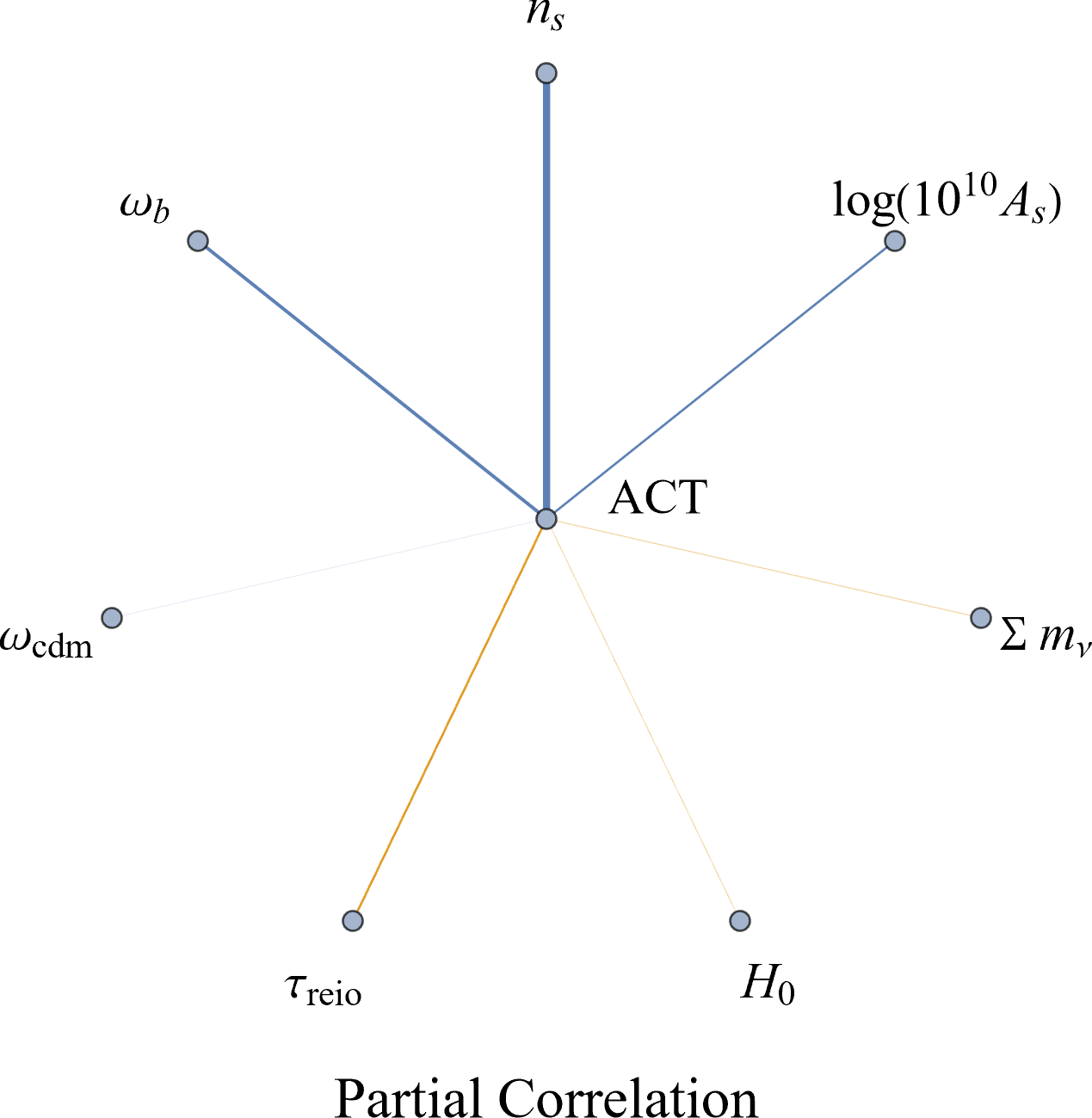}
    \caption{Correlation diagrams highlighting parameter correlations with the inclusion of ACT data. The left and right panels are the Pearson and partial correlations, respectively. The plotting style is the same as \cref{fig:corr_EECas_1}.
    }
    \label{fig:corr_ACT_1}
\end{figure}

\begin{figure}[h!]
    \centering
    \includegraphics[width=0.45\linewidth]{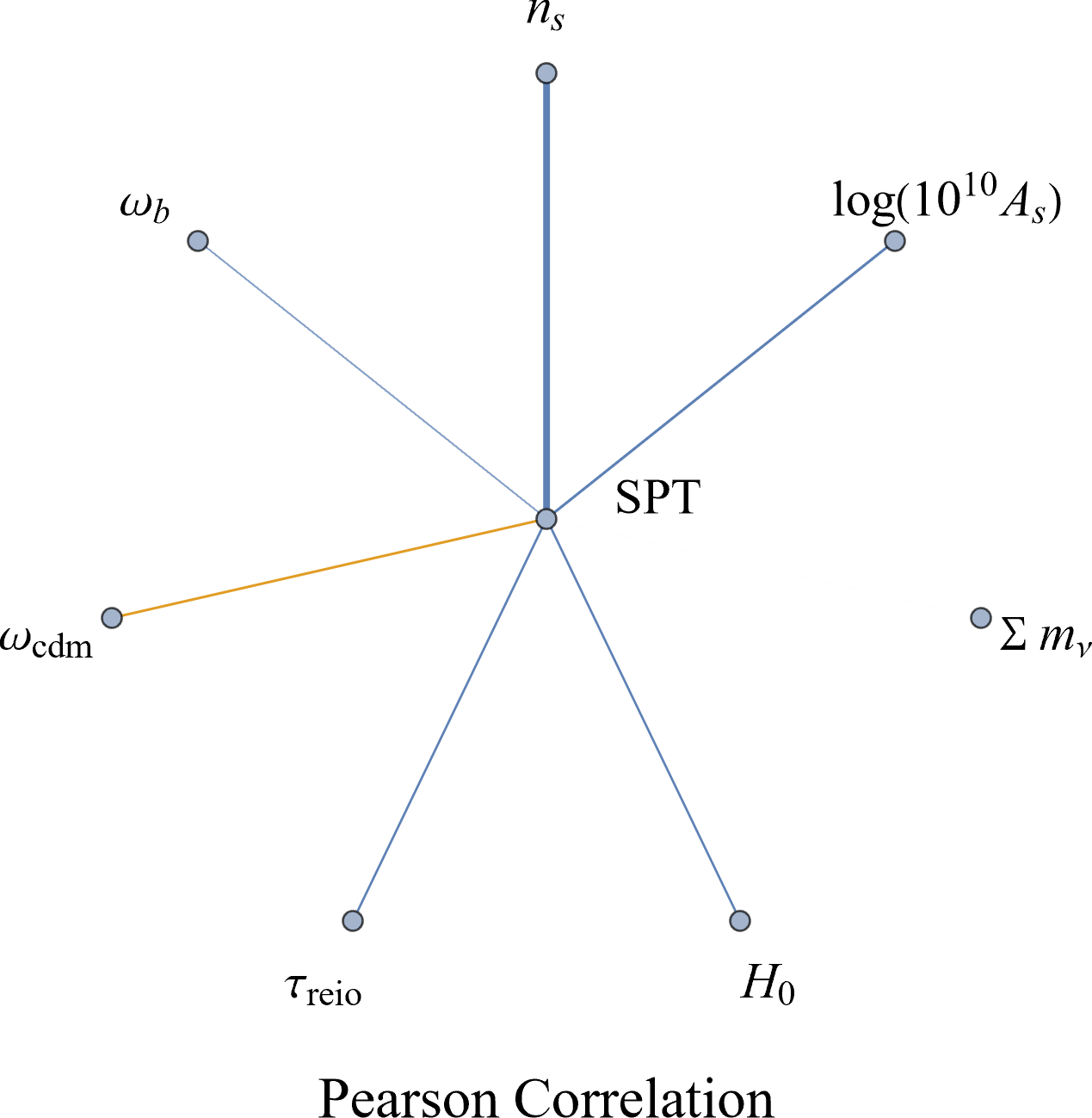}
    \includegraphics[width=0.45\linewidth]{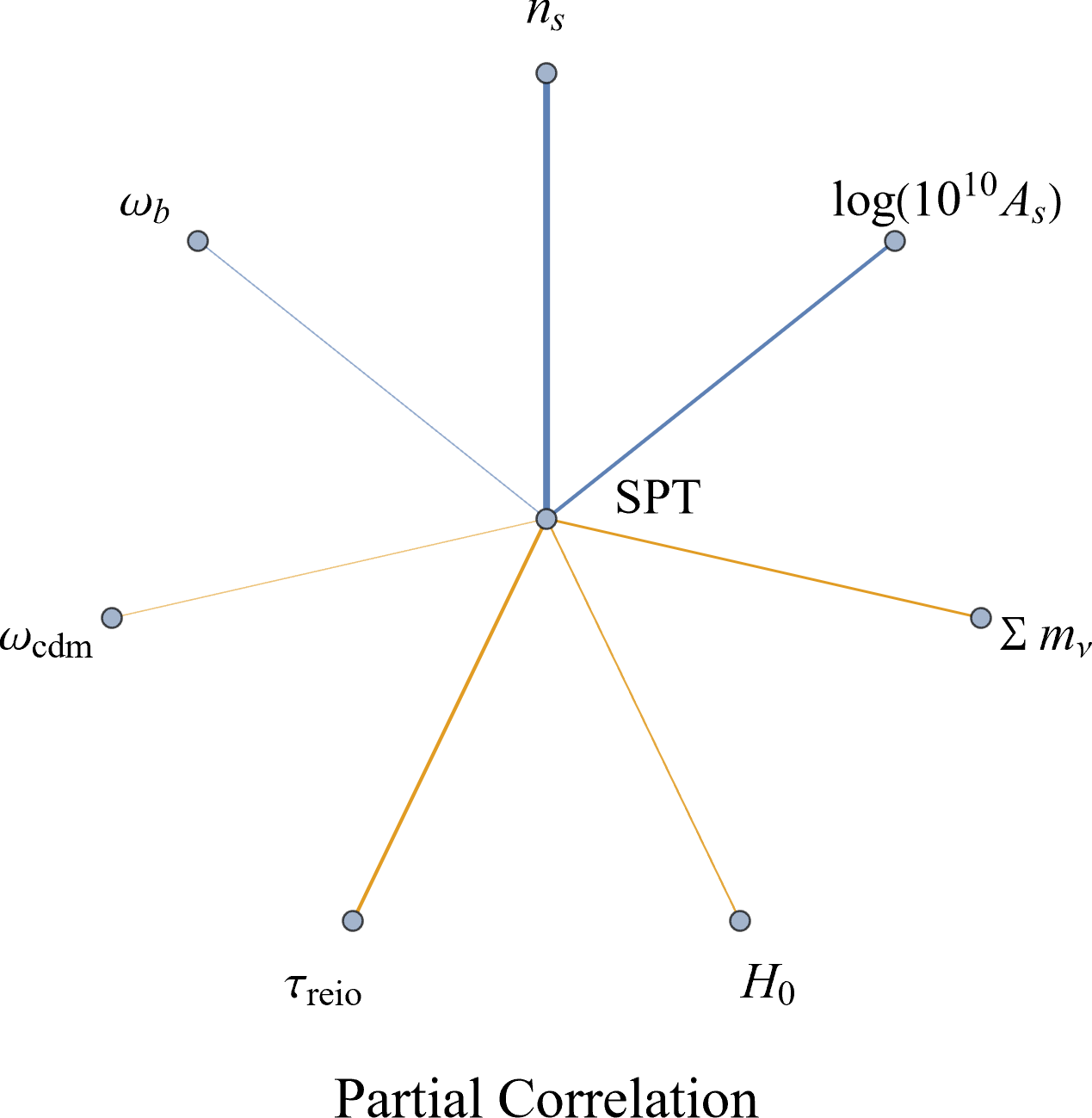}
    \caption{Correlation diagrams highlighting parameter correlations with the inclusion of SPT data. The left and right panels are the Pearson and partial correlations, respectively. The plotting style is the same as \cref{fig:corr_EECas_1}. }
    \label{fig:corr_SPT_1}
\end{figure}

Aside from the above, we also checked the correlation between the inclusion of the ACT and SPT datasets with cosmological parameters. In this case, the joint dataset is made of equal weights of \CMBP\woee+\pantheon+\bao~MC samples and either \CMBPA\woee+\pantheon+\bao~or \CMBPS\woee+\pantheon+\bao~MC samples. Each MC sample receives a binary ACT (SPT) label with value one if it comes from the \CMBPA(\CMBPS) dataset, and zero if not. The Pearson and partial correlations are shown in~\cref{fig:corr_ACT_1,fig:corr_SPT_1}. Unlike the case of the Planck low-$\ell$ EE data, these new datasets with a significant amount of information in the high-$\ell$ modes affect almost all cosmological parameters. Most elements are non-negligible in the views of both Pearson and partial correlations. Moreover, the effects of the ACT and SPT data are similar to each other in both Pearson and partial correlations, strongly indicating the consistency of the high-$\ell$ data impact. Both datasets tend to increase the value of $n_s$. The increase of $n_s$ remains obvious in terms of partial correlation, where the effects from other cosmological parameters are excluded. Similar positive correlations also appear in $A_s$ and $\omega_b$. Situations for the remaining four parameters are less straightforward, as the impact in terms of Pearson and partial correlations are often of the opposite sign. 

Finally, we also investigate briefly the affects of different supernovae datasets, \pantheon, \DES, and \union, in particular in the dynamical DE context. The cosmological parameter correlations remain qualitatively similar to the \pantheon~MC samples presented in \cref{sec:w0wa} (see \cref{app:DES_union} for the correlation coefficients for \DES~and \union.)

\section{Conclusion}
\label{sec:conclusion}

A nontrivial relationship between constraints on the optical depth of reionization $\taureio$ from CMB data and various tensions and anomalies in cosmology has begun to be uncovered. In our own recent work \cite{Allali:2025wwi} and that of others \cite{Jhaveri:2025neg,Sailer:2025lxj}, a potential increase in the inferred value of $\taureio$ has been shown to alleviate tensions between several sets of data in cosmology. In this work, we have explored discrepancies between (i) CMB and supernovae determinations of $H_0$, (ii) CMB and BAO constraints on $\Omegam$ and $h r_d$, (iii) cosmological upper bounds and experimental lower bounds on $\sum m_\nu$, and (iv) inferences of dynamical dark energy or a cosmological constant. We have summarized the behavior that follows from larger $\taureio$ in each context and taken a further step to investigate the parameter relationships responsible for these shifts in tension.

By analyzing the correlations of different sets of cosmological parameters within each model and dataset of interest, we have been able to further uncover the nature of these relationships. The Pearson correlation coefficients in \cref{tab:LCDMcorr,tab:hrdcorr,tab:mnucorr,tab:w0wacorr} clearly display that, for each tension/anomaly, $\taureio$ is correlated with the appropriate parameters such that an increase in $\taureio$ would relax the tension. In contrast, the partial correlation coefficients in the same tables demonstrate that each correlation with $\taureio$ is not intrinsic to that parameter, as the values of the partial correlations of interest tend to be close to zero. Rather, our analysis has shown that the significant Pearson correlations with $\taureio$ are a consequence of several many-dimensional parameter relationships. We can therefore conclude that, while a potential increase in $\taureio$ could relax and even relieve some of the tensions we discussed, this behavior is not a property of $\taureio$ itself.

It is interesting also to consider the likelihood of such a shift in the observed value of $\taureio$. In this work, we have periodically excluded the low-$\ell$ EE dataset from the {\it Planck} CMB data as a proxy for inferring larger $\taureio$. However, an important question remains of whether such a shift in $\taureio$ is likely to happen. At this point, only the low-$\ell$ data from {\it Planck} serves to break the parameter degeneracy between the primordial spectrum amplitude $A_s$ and $\taureio$, and thus this one set of data from a single experiment provides the most up-to-date CMB constraints on $\taureio$. There is some effort to provide complementary observations, with some progress already made by ground-based experiments such as CLASS \cite{CLASS:2025khf}, AliCPT~\cite{Li:2017drr} and future observations to be made by satellite telescopes such as {\it LiteBIRD} \cite{Paoletti:2022kij}. These will serve as independent checks on the CMB spectra at the largest scales, providing more input to the contstraints on $\taureio$. As of now, there are no significant reasons to doubt the constraints from {\it Planck} (see e.g. \cite{Ilic:2025idl} for a recent analysis using the latest data from {\it Planck} PR4). However, the low-$\ell$ EE part of the {\it Planck} spectra has the largest statistical uncertainties. Future data will improve on the precision and provide more robust constraints on $\taureio$. Thus, it may be interesting to see in the coming years whether a shift in $\taureio$ is observed, and how it would impact the biggest puzzles in cosmology.

\section*{Acknowledgements}
We would like to thank Elisa Ferreira and Joel Meyers for helpful discussions. IJA, JF and PS are supported by the NASA grant 80NSSC22K081 and the DOE grant DE-SC-0010010. This work was conducted using computational resources and services at the Center for Computation and Visualization, Brown University.

\bibliographystyle{JHEP}
\bibliography{Ref}

\pagebreak

\appendix

\section{Correlations for \CMBPAS\woee}
\label{app:PAS_corr}

We present tables of Pearson and partial correlation coefficients complementary to \cref{tab:LCDMcorr,tab:hrdcorr,tab:mnucorr}, using instead the \CMBPAS\woee+\bao+\pantheon~data combination. \cref{tab:LCDM_PAS_corr} gives correlations for the six $\Lambda$CDM parameters; \cref{tab:LCDM_PAS_hrd_corr} shows the parameter set $\{\logA,n_s,hr_d,$ $ \omega_b, \Omegam,\taureio\}$ relevant for the BAO-CMB tension discussion in \cref{subsec: bao-cmb}; and \cref{tab:LCDM_mnu_PAS_corr} gives the correlations for the seven parameters of $\Lambda$CDM+$\sum m_\nu$. The correlations are qualitatively similar to those given in \cref{tab:LCDMcorr,tab:hrdcorr,tab:mnucorr}, with the greater fluctuations in the partial correlations compared to the Pearson correlations. In addition, correlations of $n_s$ are slightly altered due to the greater sensitivity to the small-scale power spectrum in ACT/SPT data, especially the correlation between $n_s$ and $\omega_b$ which nearly switches direction when replacing \CMBP~with \CMBPAS.

\begin{table}[]
    \centering
\begin{tabular}{lcccccc}
\toprule
{\bf Pearson }& $\logA $ & $n_s$ & $H_0$ & $\omega_b$ & $\omegacdm$ & $\taureio$ \\
\midrule
$\logA$ & 1.00 & 0.30 & 0.45 & 0.20 & -0.44 & 0.97 \\
$n_s$ & 0.30 & 1.00 & 0.44 & -0.02 & -0.49 & 0.35 \\
$H_0$ & 0.45 & 0.44 & 1.00 & 0.49 & -0.91 & 0.46 \\
$\omega_b$ & 0.20 & -0.02 & 0.49 & 1.00 & -0.19 & 0.19 \\
$\omegacdm$ & -0.44 & -0.49 & -0.91 & -0.19 & 1.00 & -0.45 \\
$\taureio$ & 0.97 & 0.35 & 0.46 & 0.19 & -0.45 & 1.00 \\
\bottomrule
\end{tabular}

\begin{tabular}{lcccccc}
\toprule
{\bf Partial} & $\logA $ & $n_s$ & $H_0$ & $\omega_b$ & $\omegacdm$ & $\taureio$ \\
\midrule
$\logA$ & --- & -0.26 & 0.09 & -0.04 & 0.05 & 0.97 \\
$n_s$ & -0.26 & --- & 0.16 & -0.21 & 0.00 & 0.29 \\
$H_0$ & 0.09 & 0.16 & --- & 0.78 & -0.94 & -0.08 \\
$\omega_b$ & -0.04 & -0.21 & 0.78 & --- & 0.70 & 0.06 \\
$\omegacdm$ & 0.05 & 0.00 & -0.94 & 0.70 & --- & -0.07 \\
$\taureio$ & 0.97 & 0.29 & -0.08 & 0.06 & -0.07 & --- \\
\bottomrule
\end{tabular}
    \caption{Pearson correlations (top) and partial correlations (bottom) between the standard six parameters in $\Lambda$CDM  with the dataset \CMBPAS\woee+\bao+\pantheon.}
    \label{tab:LCDM_PAS_corr}
\end{table}

\begin{table}[]
    \centering
\begin{tabular}{lcccccc}
\toprule
 {\bf Pearson} & $\logA $ & $n_s$ & $h r_d \, [\mathrm{Mpc}]$ & $\omega_b$ & $\Omegam$ & $\taureio$ \\
\midrule
$\logA$ & 1.00 & 0.30 & 0.45 & 0.20 & -0.46 & 0.97 \\
$n_s$ & 0.30 & 1.00 & 0.49 & -0.02 & -0.48 & 0.35 \\
$h r_d \, [\mathrm{Mpc}]$ & 0.45 & 0.49 & 1.00 & 0.30 & -1.00 & 0.46 \\
$\omega_b$ & 0.20 & -0.02 & 0.30 & 1.00 & -0.34 & 0.19 \\
$\Omegam$ & -0.46 & -0.48 & -1.00 & -0.34 & 1.00 & -0.46 \\
$\taureio$ & 0.97 & 0.35 & 0.46 & 0.19 & -0.46 & 1.00 \\
\bottomrule
\end{tabular}

\begin{tabular}{lcccccc}
\toprule
 {\bf Partial} & $\logA $ & $n_s$ & $h r_d \, [\mathrm{Mpc}]$ & $\omega_b$ & $\Omegam$ & $\taureio$ \\
\midrule
$\logA$ & --- & -0.26 & 0.06 & 0.06 & 0.06 & 0.97 \\
$n_s$ & -0.26 & --- & 0.07 & -0.12 & 0.04 & 0.29 \\
$h r_d \, [\mathrm{Mpc}]$ & 0.06 & 0.07 & --- & -0.67 & -1.00 & -0.07 \\
$\omega_b$ & 0.06 & -0.12 & -0.67 & --- & -0.68 & -0.05 \\
$\Omegam$ & 0.06 & 0.04 & -1.00 & -0.68 & --- & -0.06 \\
$\taureio$ & 0.97 & 0.29 & -0.07 & -0.05 & -0.06 & --- \\
\bottomrule
\end{tabular}
    \caption{Pearson correlations (top) and partial correlations (bottom) between the parameters in the set $\{\logA,n_s,hr_d,\omega_b,\Omegam,\taureio\}$ in $\Lambda$CDM  with the dataset \CMBPAS\woee+\bao+\pantheon.}
    \label{tab:LCDM_PAS_hrd_corr}
\end{table}

\begin{table}[]
    \centering
\begin{tabular}{lccccccc}
\toprule
{\bf Pearson} & $\logA$ & $n_s$ & $H_0$ & $\omega_b$ & $\omegacdm$ & $\taureio$ & $\sum m_\nu$ \\
\midrule
$\logA$ & 1.00 & 0.50 & -0.11 & 0.27 & -0.69 & 0.98 & 0.68 \\
$n_s$ & 0.50 & 1.00 & 0.10 & 0.06 & -0.64 & 0.53 & 0.44 \\
$H_0$ & -0.11 & 0.10 & 1.00 & 0.28 & -0.21 & -0.08 & -0.55 \\
$\omega_b$ & 0.27 & 0.06 & 0.28 & 1.00 & -0.27 & 0.25 & 0.20 \\
$\omegacdm$ & -0.69 & -0.64 & -0.21 & -0.27 & 1.00 & -0.69 & -0.65 \\
$\taureio$ & 0.98 & 0.53 & -0.08 & 0.25 & -0.69 & 1.00 & 0.65 \\
$\sum m_\nu$ & 0.68 & 0.44 & -0.55 & 0.20 & -0.65 & 0.65 & 1.00 \\
\bottomrule
\end{tabular}

\begin{tabular}{lccccccc}
\toprule
{\bf Partial} & $\logA$ & $n_s$ & $H_0$ & $\omega_b$ & $\omegacdm$ & $\taureio$ & $\sum m_\nu$ \\
\midrule
$\logA$ & --- & -0.25 & 0.11 & -0.06 & 0.07 & 0.97 & 0.17 \\
$n_s$ & -0.25 & --- & 0.13 & -0.19 & -0.04 & 0.28 & 0.14 \\
$H_0$ & 0.11 & 0.13 & --- & 0.79 & -0.93 & -0.10 & -0.97 \\
$\omega_b$ & -0.06 & -0.19 & 0.79 & --- & 0.69 & 0.07 & 0.76 \\
$\omegacdm$ & 0.07 & -0.04 & -0.93 & 0.69 & --- & -0.09 & -0.93 \\
$\taureio$ & 0.97 & 0.28 & -0.10 & 0.07 & -0.09 & --- & -0.15 \\
$\sum m_\nu$ & 0.17 & 0.14 & -0.97 & 0.76 & -0.93 & -0.15 & --- \\
\bottomrule
\end{tabular}
    \caption{Pearson correlations (top) and partial correlations (bottom) between the seven parameters in $\Lambda$CDM+$\sum m_\nu$  with the dataset \CMBPAS\woee+\bao+\pantheon.}
    \label{tab:LCDM_mnu_PAS_corr}
\end{table}

\pagebreak

\section{Neutrino Mass with \DES}
\label{app:nu_mass_DES}

\begin{figure}
    \centering
    \includegraphics[width=0.73\linewidth]{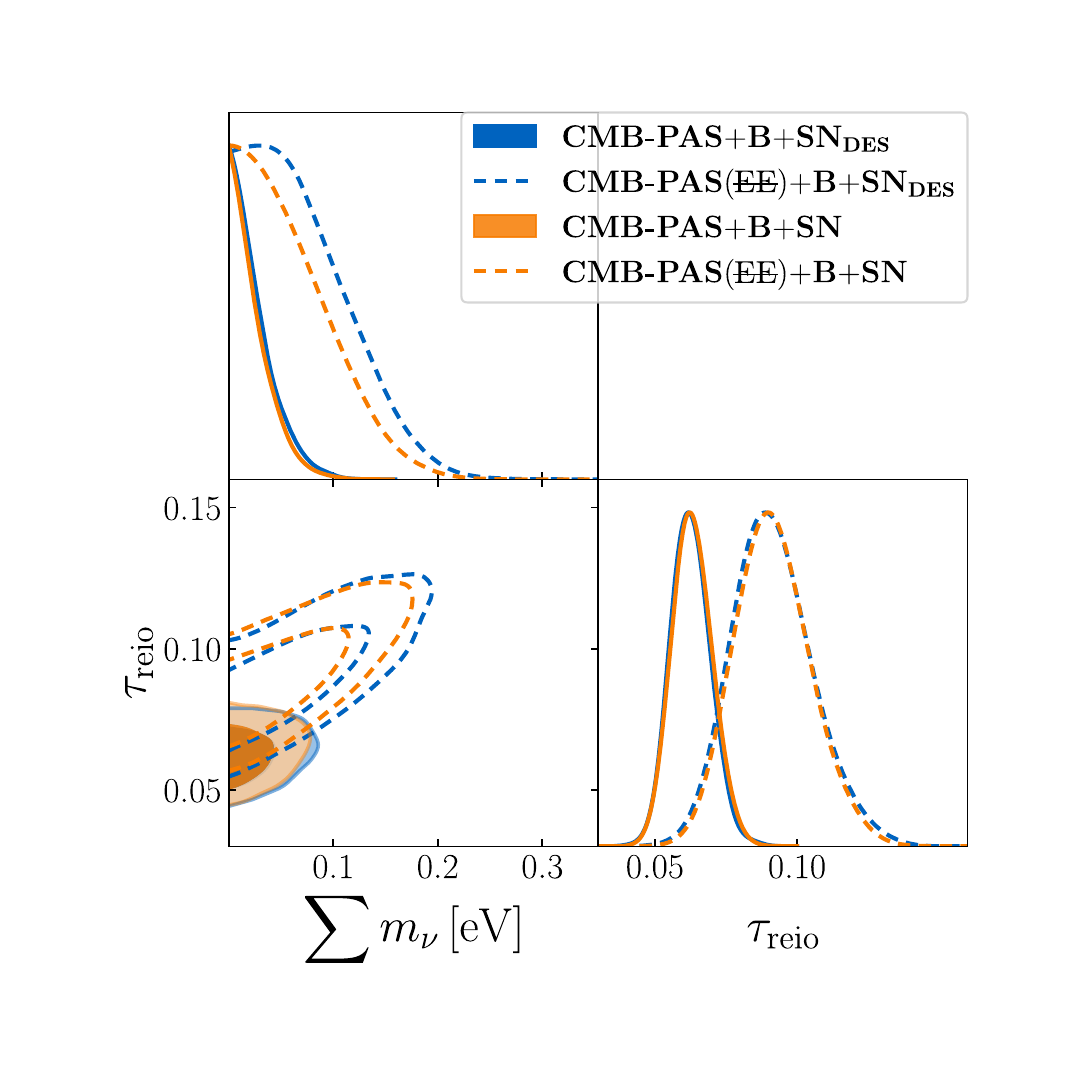}
    \caption{The one- and two-dimensional posterior distributions of $\taureio$ and $\sum m_\nu$ in a series of fits to four different combinations of datasets, assuming the $\Lambda$CDM+$\sum m_\nu$ model.  }
    \label{fig:mnu_tau_DES}
\end{figure}

As pointed out in \cref{sec:nu_mass}, the combination of {\it Planck}, ACT, and SPT CMB data, in the absence of the low-$\ell$ EE data, begins to show a peaked posterior for $\sum m_\nu$. In the case of the data combination \CMBPAS\woee+\bao+\pantheon, this peak is very marginally present. However, as shown in \cite{Allali:2024aiv,Loverde:2024nfi}, constraints on neutrino masses depend significantly on the choice of supernovae data. As an example, we present in \cref{fig:mnu_tau_DES} also a fit using the DES supernovae dataset. One can see that the neutrino mass constraint is most significantly relaxed with the \CMBPAS\woee+\bao+\DES~dataset, corresponding to $\sum m_\nu < 0.16$ eV at 95\% C.L., and indeed this posterior even shows the clear formation of a peak away from zero. In such a context, then, the observation of ``negative neutrino mass" is no longer present.

\section{Correlations for \DES~and \union~in $w_0w_a$CDM}\label{app:DES_union}

\begin{figure}
    \centering
    \includegraphics[width=0.6\linewidth]{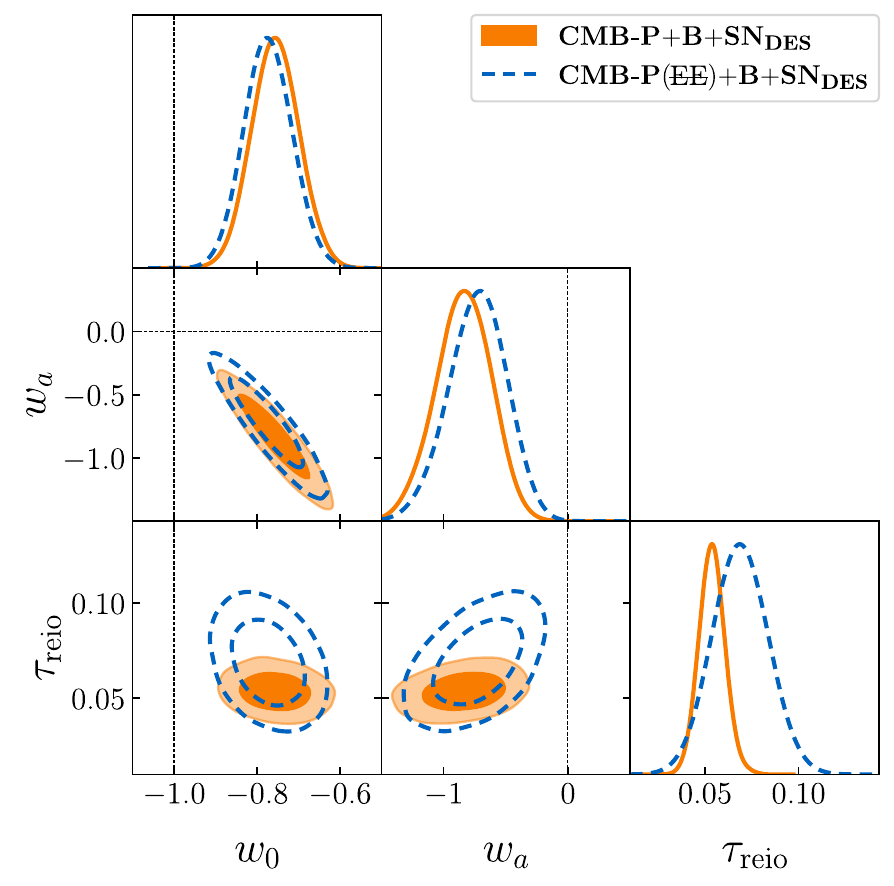}
    \caption{The one- and two-dimensional posterior distributions of $\taureio$, $w_0$, and $w_a$ when fit to \CMBP+\bao+\DES~and \CMBP\woee+\bao+\DES.}
    \label{fig:w0wa_DES}
\end{figure}

\begin{figure}
    \centering
    \includegraphics[width=0.6\linewidth]{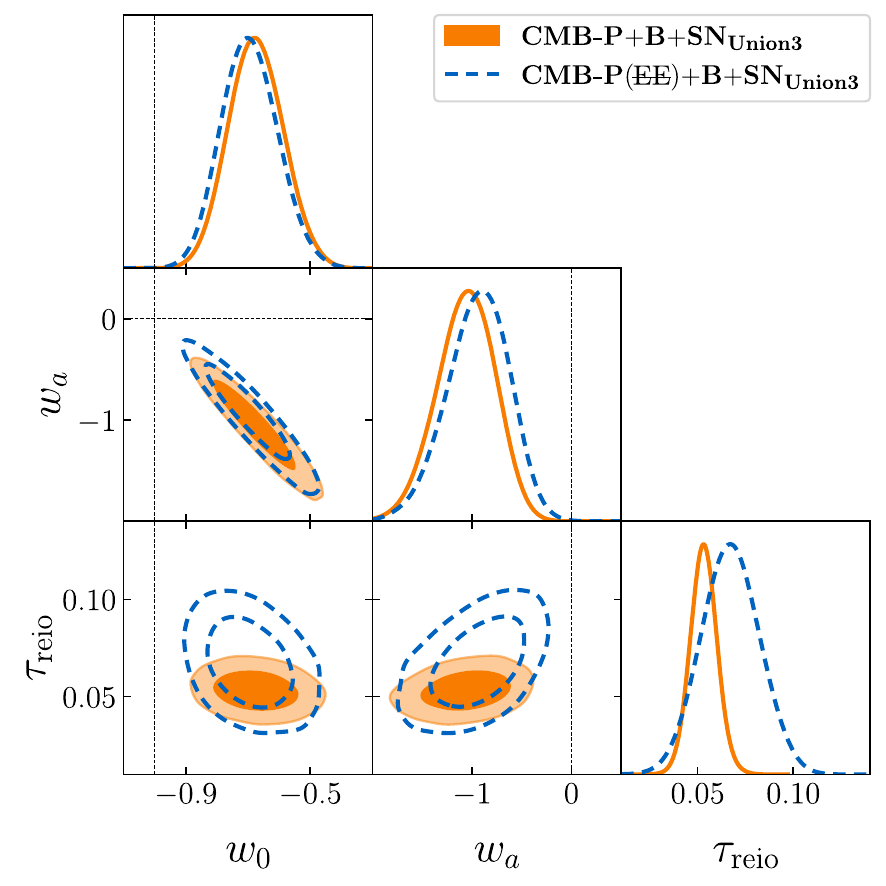}
    \caption{The one- and two-dimensional posterior distributions of $\taureio$, $w_0$, and $w_a$ when fit to \CMBP+\bao+\union~and \CMBP\woee+\bao+\union.}
    \label{fig:w0wa_union}
\end{figure}

We provide tables of the Pearson and partial correlation coefficients for the $w_0w_a$CDM model parameters when fitting to \CMBP\woee+\bao+\DES~in \cref{tab:w0wa_DES_corr} and \CMBP\woee+\bao+ \union~in \cref{tab:w0wa_union_corr}, analogous to \cref{tab:w0wacorr} displaying the case of \CMBP\woee+\bao+\pantheon. The correlations are nearly identical, with minor fluctuations, for any given choice of supernovae dataset. \cref{fig:w0wa_DES,fig:w0wa_union} display posterior distributions when fitting to these datasets for comparison with \cref{fig:w0wa_tau}.

\begin{table}[]
    \centering
    \begin{tabular}{lcccccccc}
\toprule
{\bf Pearson} & $\logA $ & $n_s$ & $H_0$ & $\omega_b$ & $\omegacdm$ & $\taureio$ & $w_0$ & $w_a$ \\
\midrule
$\logA$ & 1.00 & 0.51 & -0.01 & 0.44 & -0.65 & 0.99 & -0.31 & 0.49 \\
$n_s$ & 0.51 & 1.00 & 0.00 & 0.44 & -0.67 & 0.52 & -0.26 & 0.44 \\
$H_0$ & -0.01 & 0.00 & 1.00 & 0.08 & 0.01 & -0.01 & -0.53 & 0.18 \\
$\omega_b$ & 0.44 & 0.44 & 0.08 & 1.00 & -0.52 & 0.43 & -0.20 & 0.37 \\
$\omegacdm$ & -0.65 & -0.67 & 0.01 & -0.52 & 1.00 & -0.65 & 0.39 & -0.64 \\
$\taureio$ & 0.99 & 0.52 & -0.01 & 0.43 & -0.65 & 1.00 & -0.30 & 0.48 \\
$w_0$ & -0.31 & -0.26 & -0.53 & -0.20 & 0.39 & -0.30 & 1.00 & -0.90 \\
$w_a$ & 0.49 & 0.44 & 0.18 & 0.37 & -0.64 & 0.48 & -0.90 & 1.00 \\
\bottomrule
\end{tabular}

\begin{tabular}{lcccccccc}
\toprule
 {\bf Partial} & $\logA $ & $n_s$ & $H_0$ & $\omega_b$ & $\omegacdm$ & $\taureio$ & $w_0$ & $w_a$ \\
\midrule
$\logA$ & --- & -0.10 & 0.05 & 0.04 & 0.06 & 0.98 & 0.05 & 0.06 \\
$n_s$ & -0.10 & --- & 0.05 & 0.05 & -0.12 & 0.12 & 0.05 & 0.05 \\
$H_0$ & 0.05 & 0.05 & --- & 0.76 & -0.89 & -0.04 & -0.99 & -0.98 \\
$\omega_b$ & 0.04 & 0.05 & 0.76 & --- & 0.63 & -0.03 & 0.76 & 0.75 \\
$\omegacdm$ & 0.06 & -0.12 & -0.89 & 0.63 & --- & -0.08 & -0.91 & -0.92 \\
$\taureio$ & 0.98 & 0.12 & -0.04 & -0.03 & -0.08 & --- & -0.05 & -0.05 \\
$w_0$ & 0.05 & 0.05 & -0.99 & 0.76 & -0.91 & -0.05 & --- & -1.00 \\
$w_a$ & 0.06 & 0.05 & -0.98 & 0.75 & -0.92 & -0.05 & -1.00 & --- \\
\bottomrule
\end{tabular}
    \caption{Pearson correlations (top) and partial correlations (bottom) between the cosmological parameters in the $w_0w_a$CDM model with the dataset \CMBP\woee+\bao+\DES.}
    \label{tab:w0wa_DES_corr}
\end{table}

\begin{table}[]
    \centering
    \begin{tabular}{lcccccccc}
\toprule
 {\bf Pearson} & $\logA $ & $n_s$ & $H_0$ & $\omega_b$ & $\omegacdm$ & $\taureio$ & $w_0$ & $w_a$ \\
\midrule
$\logA$ & 1.00 & 0.51 & 0.10 & 0.45 & -0.66 & 0.99 & -0.32 & 0.48 \\
$n_s$ & 0.51 & 1.00 & 0.09 & 0.44 & -0.67 & 0.52 & -0.28 & 0.43 \\
$H_0$ & 0.10 & 0.09 & 1.00 & 0.14 & -0.12 & 0.10 & -0.78 & 0.53 \\
$\omega_b$ & 0.45 & 0.44 & 0.14 & 1.00 & -0.52 & 0.43 & -0.24 & 0.37 \\
$\omegacdm$ & -0.66 & -0.67 & -0.12 & -0.52 & 1.00 & -0.66 & 0.41 & -0.63 \\
$\taureio$ & 0.99 & 0.52 & 0.10 & 0.43 & -0.66 & 1.00 & -0.32 & 0.47 \\
$w_0$ & -0.32 & -0.28 & -0.78 & -0.24 & 0.41 & -0.32 & 1.00 & -0.93 \\
$w_a$ & 0.48 & 0.43 & 0.53 & 0.37 & -0.63 & 0.47 & -0.93 & 1.00 \\
\bottomrule
\end{tabular}

\begin{tabular}{lcccccccc}
\toprule
{\bf Partial} & $\logA $ & $n_s$ & $H_0$ & $\omega_b$ & $\omegacdm$ & $\taureio$ & $w_0$ & $w_a$ \\
\midrule
$\logA$ & --- & -0.10 & 0.03 & 0.07 & 0.05 & 0.98 & 0.04 & 0.05 \\
$n_s$ & -0.10 & --- & 0.08 & 0.03 & -0.10 & 0.13 & 0.08 & 0.08 \\
$H_0$ & 0.03 & 0.08 & --- & 0.75 & -0.88 & -0.03 & -0.99 & -0.98 \\
$\omega_b$ & 0.07 & 0.03 & 0.75 & --- & 0.61 & -0.05 & 0.75 & 0.75 \\
$\omegacdm$ & 0.05 & -0.10 & -0.88 & 0.61 & --- & -0.08 & -0.90 & -0.91 \\
$\taureio$ & 0.98 & 0.13 & -0.03 & -0.05 & -0.08 & --- & -0.04 & -0.05 \\
$w_0$ & 0.04 & 0.08 & -0.99 & 0.75 & -0.90 & -0.04 & --- & -1.00 \\
$w_a$ & 0.05 & 0.08 & -0.98 & 0.75 & -0.91 & -0.05 & -1.00 & --- \\
\bottomrule
\end{tabular}
    \caption{Pearson correlations (top) and partial correlations (bottom) between the cosmological parameters in the $w_0w_a$CDM model with the dataset \CMBP\woee+\bao+\union.}
    \label{tab:w0wa_union_corr}
\end{table}

\section{Posteriors for \CMBPAS}\label{app:PAS_post}

\begin{figure}
    \centering
    \includegraphics[width=1.0\linewidth]{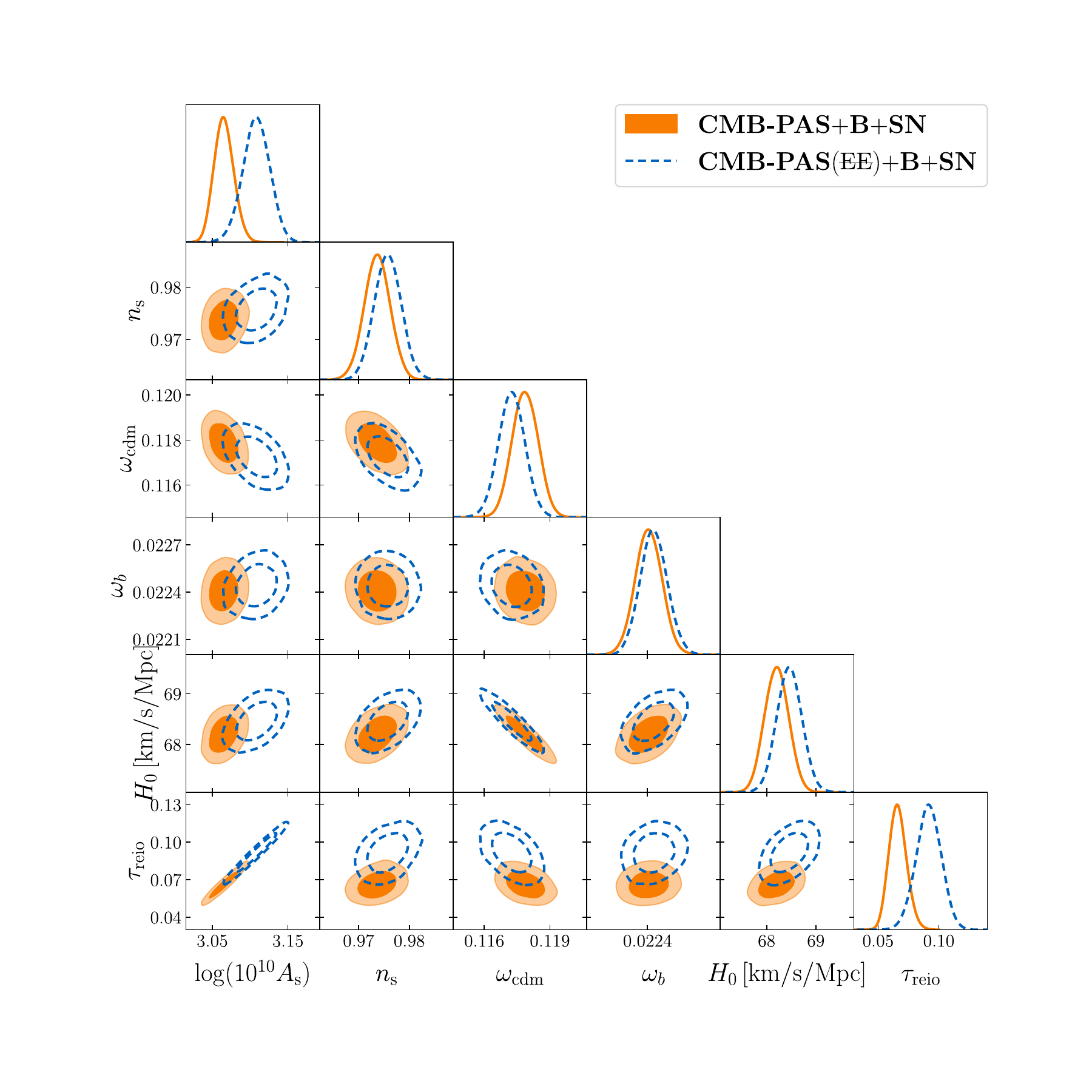}
    \caption{The one- and two-dimensional posterior distributions of all cosmological parameters for the $\Lambda$CDM model when fit to \CMBPAS+\bao+\pantheon~and \CMBPAS\woee+\bao+\pantheon.}
    \label{fig:PAS_full}
\end{figure}

\begin{figure}
    \centering
    \includegraphics[width=1.0\linewidth]{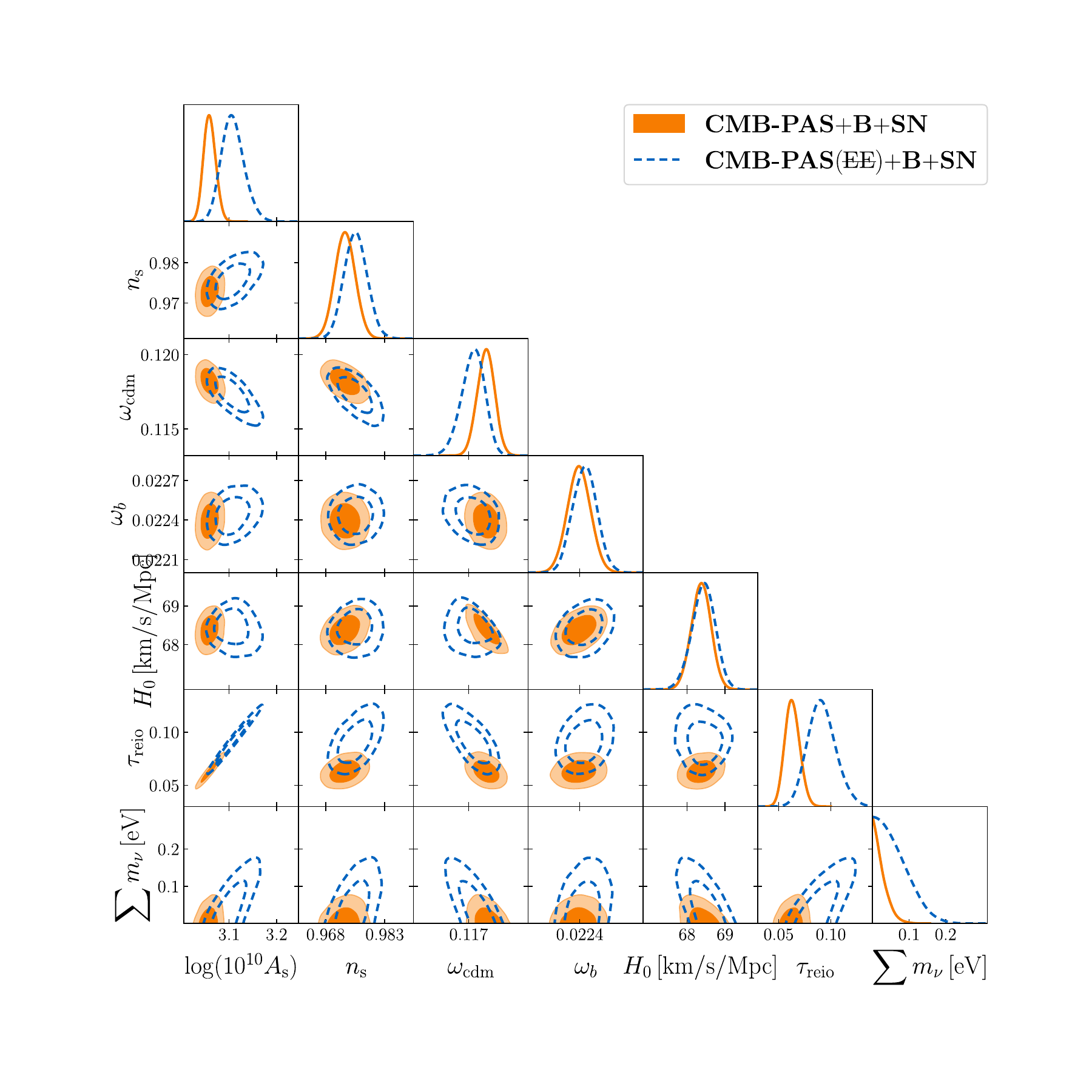}
    \caption{The one- and two-dimensional posterior distributions of all cosmological parameters for the $\Lambda$CDM+$\sum m_\nu$ model when fit to \CMBPAS+\bao+\pantheon~and \CMBPAS\woee+\bao+\pantheon.}
    \label{fig:PAS_full_mnu}
\end{figure}

The combination \CMBPAS~combines CMB data from {\it Planck}, ACT, and SPT, using the highest precision from each dataset to try to get the most constraining power from CMB data. We present posterior statistics and plots from this novel combination in this section. \cref{tab:LCDM_PAS_stats} gives the 68\% C.L. marginalized statistics for $\Lambda$CDM parameters fit to the combinations \CMBPAS+\bao+\pantheon~and \CMBPAS\woee+\bao+\pantheon, with the corresponding posterior distributions plotted in \cref{fig:PAS_full}. In addition, posteriors for $\Lambda$CDM+$\sum m_\nu$ are given in \cref{tab:LCDM_PAS_mnu_stats,fig:PAS_full_mnu}. 

We obtain a novel upper bound on the total neutrino mass $\sum m_\nu < 0.061$ eV, which is slightly weaker than the bound from the combination of {\it Planck}, ACT, and SPT implemented by the SPT collaboration in \cite{SPT-3G:2025bzu} ($\sum m_\nu < 0.048$ eV). This difference is most likely driven by our inclusion of the Pantheon+ supernovae dataset, which is known to increase the bound on $\sum m_\nu$ (see \cite{Allali:2024aiv,Loverde:2024nfi}), but it may also be somewhat dependent on our different choices of $\ell$-ranges for each CMB datset compared to the choices made by the SPT collaboration. In addition, the prior on $\taureio$ utilized by the SPT collaboration has also been shown to drive down the constraint a bit, see \cite{Graham:2025fdt} for more details. Nonetheless, since this upper bound includes all current CMB datasets, the most recent DESI BAO results, and one of three complementary supernovae datasets, it can be considered the most complete and up-to-date constraint on neutrino mass in the $\Lambda$CDM+$\sum m_\nu$ context.

\begin{table*}

\centering

\begin{tabular} {| l | c| c|}
\hline\hline
 \multicolumn{1}{|c|}{ Parameter} &  \multicolumn{1}{|c|}{\CMBPAS+\bao+\pantheon} &  \multicolumn{1}{|c|}{\CMBPAS\woee+\bao+\pantheon}\\
\hline\hline
$\logA$ & $3.065^{+0.012}_{-0.014}   $ & $3.108^{+0.018}_{-0.018}   $\\
$n_s              $ & $0.9737^{+0.0025}_{-0.0025}$ & $0.9758^{+0.0026}_{-0.0026}$\\
$\omegacdm      $ & $0.11787^{+0.00059}_{-0.00059}$ & $0.11725^{+0.00060}_{-0.00059}$\\
$\omega_b $ & $0.022408^{+0.000087}_{-0.000086}$ & $0.022442^{+0.000088}_{-0.000088}$\\
$H_0$   [km/s/Mpc]                   & $68.20^{+0.24}_{-0.24}     $ & $68.46^{+0.25}_{-0.25}     $\\
$\taureio       $ & $0.0663^{+0.0066}_{-0.0078}$ & $0.092^{+0.010}_{-0.010}   $\\
\hline
\end{tabular}

\caption{Posterior mean values and corresponding 68\% C.L. intervals for the cosmological parameters in the $\Lambda$CDM model when fit to \CMBPAS+\bao+\pantheon~and \CMBPAS\woee+\bao+\pantheon.}
\label{tab:LCDM_PAS_stats}

\end{table*}

\begin{table*}

\centering

\begin{tabular} {| l | c| c|}
\hline\hline
 \multicolumn{1}{|c|}{ Parameter} &  \multicolumn{1}{|c|}{\CMBPAS+\bao+\pantheon} &  \multicolumn{1}{|c|}{\CMBPAS\woee+\bao+\pantheon}\\
\hline\hline
$\logA$ & $3.059^{+0.012}_{-0.014}   $ & $3.108^{+0.021}_{-0.026}   $\\
$n_{s}              $ & $0.9729^{+0.0025}_{-0.0026}$ & $0.9756^{+0.0029}_{-0.0029}$\\
$\omegacdm      $ & $0.11819^{+0.00059}_{-0.00059}$ & $0.11728^{+0.00087}_{-0.00071}$\\
$\omega_b      $ & $0.022394^{+0.000089}_{-0.000089}$ & $0.022439^{+0.000092}_{-0.000095}$\\
$H_0$  [km/s/Mpc]                    & $68.37^{+0.25}_{-0.26}     $ & $68.43^{+0.32}_{-0.29}     $\\
$\taureio        $ & $0.0631^{+0.0067}_{-0.0076}$ & $0.091^{+0.012}_{-0.015}   $\\
$\sum m_\nu$ [eV]              & $< 0.0613                  $ (95\% C.L.) & $< 0.142                  $ (95\% C.L.)\\
\hline
\end{tabular}

\caption{Posterior mean values and corresponding 68\% C.L. intervals for the cosmological parameters in the $\Lambda$CDM+$\sum m_\nu$ model when fit to \CMBPAS+\bao+\pantheon~and \CMBPAS\woee+\bao+\pantheon.}
\label{tab:LCDM_PAS_mnu_stats}
\end{table*}

\end{document}